\documentclass[12pt,draftclsnofoot,onecolumn]{IEEEtran}
\usepackage[utf8]{inputenc} 
\usepackage[T1]{fontenc}
\usepackage{url}
\usepackage{ifthen}
\usepackage{cite}
\usepackage[cmex10]{amsmath}
\allowdisplaybreaks[4]
\newcommand{\Mod}[1]{\ (\mathrm{mod}\ #1)}
\usepackage{graphicx}
\usepackage{amsfonts}
\usepackage{enumitem}
\usepackage{amssymb}
\usepackage{mathrsfs}
\usepackage{bm}
\usepackage{mdwmath}
\usepackage{subfigure}
\usepackage{mdwtab}
\usepackage{dsfont}
\usepackage{booktabs}
\usepackage[ruled,vlined,linesnumbered]{algorithm2e} 
\usepackage{threeparttable}
\usepackage{tcolorbox}

\newtheorem{theorem}{Theorem}

\newtheorem{prop}{Proposition}
\newtheorem{defi}{Definition}

\newenvironment{iarray}{\begin{IEEEeqnarray}{rCl}}{\end{IEEEeqnarray}\ignorespacesafterend}

\newcommand{\tabincell}[2]{\begin{tabular}{@{}#1@{}}#2\end{tabular}}

\begin{document}
    \title{Age of Information Optimized MAC in V2X Sidelink via Piggyback-Based Collaboration}
    \author{Fei Peng, Zhiyuan Jiang, Shunqing Zhang, and Shugong Xu,~\emph{Fellow,~IEEE}
	\thanks{
    The authors are with Shanghai Institute for Advanced Communication and Data Science, Shanghai University, China. Emails: \{pfly\_shmily, jiangzhiyuan, shunqing, shugong\}@shu.edu.cn. The corresponding author is Zhiyuan Jiang. A preliminary version of the paper has been submitted to IEEE VTC 2020 Fall \cite{peng2020piggyback}.
    }
    }
    \maketitle
    
    \begin{abstract}
    Real-time status update in future vehicular networks is vital to enable control-level cooperative autonomous driving. Cellular Vehicle-to-Everything (C-V2X), as one of the most promising vehicular wireless technologies, adopts a Semi-Persistent Scheduling (SPS) based Medium-Access-Control (MAC) layer protocol for its sidelink communications. Despite the recent and ongoing efforts to optimize SPS, very few work has considered the status update performance of SPS. In this paper, Age of Information (AoI) is first leveraged to evaluate the MAC layer performance of C-V2X sidelink. Critical issues of SPS, i.e., persistent packet collisions and Half-Duplex (HD) effects, are identified to hinder its AoI performance. Therefore, a piggyback-based collaboration method is proposed accordingly, whereby vehicles collaborate to inform each other of potential collisions and collectively afford HD errors, while entailing only a small signaling overhead. Closed-form AoI performance is derived for the proposed scheme, optimal configurations for key parameters are hence calculated, and the convergence property is proved for decentralized implementation. Simulation results show that compared with the standardized SPS and its state-of-the-art enhancement schemes, the proposed scheme shows significantly better performance, not only in terms of AoI, but also of conventional metrics such as transmission reliability. 
    \end{abstract}
    

    \section{Introduction} 
    \label{sec_intro}     
    
    Although Vehicle-to-everything (V2X) communications are envisioned to significantly improve the reliability and efficiency of autonomous driving, their implementation is still faced with many challenges \cite{chen2017vehicle, machardy2018v2x,zhangshan20,sunyx17}. The reason behind is that, in nature, V2X is different from conventional content communications, for that it has to meet the high-reliability and low latency requirements with a large number of vehicles simultaneously, while they are travelling at high speeds. In addition, V2X technologies are preferred to be decentralized, such that vehicles can communicate with each other out of cellular coverage. Therefore, the communication protocol of V2X needs to let the terminals decide, e.g., resource allocations, considering the unique patterns of vehicular communications wherein traffic and status updates are mostly periodic.
    
    In this regard, the 3rd Generation Partnership Project (3GPP) community has standardized a semi-persistent scheduling (SPS) scheme for decentralized resource allocation in Cellular-V2X (C-V2X) mode 4. In contrast to mode 3 which involves centralized scheduling that is identical with conventional Long Term Evolution (LTE), mode 4 is fully decentralized. In a practical scenario of vehicular direct communications, distributed resource allocation is often more advantageous than centralized resource allocation from the perspective of latency and overhead. Compared with distributed resource allocation, in addition to the disadvantage of the limited coverage range for centralized resource allocation, it needs more interactions between base stations and vehicles, which may cause more latency and overhead. Furthermore, vehicles with high speed may experience frequent handover if connected to base stations. Based on these reasons, distributed resource allocation is more suitable for vehicular direct communications. 
    
    However, the reliability of current C-V2X-based distributed allocation schemes is far below that of centralized allocation and the channel utilization is low as well. More specifically, most current distributed resource allocation schemes are sensing-based only and semi-persistent. Once a collision occurs, the related vehicles could not realize the collision due to the Half-Duplex (HD) effect, referring to the fact that a terminal cannot receive and transmit simultaneously. Moreover, the characteristic of semi-persistent will result in consecutive packet collisions without being realized by the transmitters until entering the re-selection process---this is not desirable for scenarios that need high reliability and status timeliness such as autonomous driving. In addition, though the re-selection process in semi-persistent scheduling schemes can avoid persistent packet collision and is suitable in mobile scenarios to some extent, additional packet collisions may occur with a certain probability due to the re-selection process. For example, a packet collides with another packet after a re-selection process but was not collided before---this clearly needs to be avoided. Therefore, the problems mentioned above lead to an unsatisfactory performance of distributed resource allocation in V2X sidelink.
    
    In existing works and 3GPP standards, reliability and latency are often used to evaluate the performance of vehicular communications. However, the requirements of the two metrics cannot achieve a good trade-off, where the reliability performance can be enhanced at the cost of increasing the latency and vice versa. In the other word, the reliability performance with different latency cannot be compared. Therefore, it is necessary to look for a new metric to synthetically reflect the performance of reliability and latency. In addition, the two metrics cannot directly reflect the timeliness of status information, e.g., speed and locations. In this paper, Age of Information (AoI) \cite{kaul12,jiang20ai} is accordingly adopted as the performance metric that is closely related to the status freshness which is vital for vehicular communications. Formally, AoI is defined for a status and a destination, as the time elapsed since the generation of the latest status received at the destination. More specifically, AoI in this work is defined as the metric of the status freshness from any other vehicle, which will linearly increase with time during the communications. When the next packet is received, AoI will return to zero, where the time of transmission and signal process is omitted. From our perspective, AoI is the metric that can reflect a synthetic performance of reliability and latency. It is because both lower latency with the same reliability and higher reliability with the same latency can result in a lower average AoI. Based on AoI, the performance evaluation will not be restrained by the table of reliability and latency requirements and only one metric, i.e. AoI, needs to be considered. Consequently, AoI is considered as a more essential metric in the status update scenarios and AoI optimizations implicitly encompass reliability and latency enhancements.
    
    In this paper, we propose a distributed resource allocation scheme for C-V2X based on AoI optimizations. Three major features of the scheme are collaboration, sub-frame offset and persistent resource occupancy, which significantly reduce packet collisions and guarantee that each vehicle can receive messages from any other vehicle periodically. To be specific, collaboration is achieved through piggyback-like feedback from vehicles that have sensed the channels and have packets to transmit. Different from the conventional piggyback mechanisms \cite{rozner2009soar}, the piggyback in this work carries information about collision resolution messages. Therefore, collaboration in our scheme aims to directly solve the packet collisions by the receiver transmitting the collaboration message, which is fundamentally different from the existing collaboration schemes that announce self parameter configurations \cite{bonjorn2018enhanced,jeon2018reducing}. Moreover, the piggyback-based collaboration message is designed to entail only a small signaling overhead. Our main contributions include the following.
    
    \begin{itemize}
        \item 
        In vehicular direct communications, specifically C-V2X distributed Media Access Control (MAC), this is the first work to adopt and optimize AoI as the metric to enhance status update timeliness. A practical (with minimal modifications to the current standards) and effective AoI-based MAC scheme is proposed that exhibits significantly improved AoI performance. In addition, the AoI-optimized proposed scheme also shows advantages in conventional metrics such as reliability and latency. 
        \item 
        The convergence of the proposed scheme is proved; theoretical performance analysis in the scenarios of both static and dynamic vehicular traffic flows is presented. Shown by the theoretical results, when the Channel Busy Ratio (CBR) is in a certain range, the average AoI of all vehicles by our proposed scheme is independent with the vehicle population in the static scenario and is only slightly affected in the dynamic scenario---this proposition demonstrates the scalability of the proposed scheme. The optimal Resource Reservation Interval (RRI) that minimizes AoI is also derived and validated by analysis and simulations.
        \item 
        Extensive computer simulations are respectively conducted in the MAC layer and the freeway scenario. From the perspective of the MAC layer, compared with the current C-V2X distributed resource allocation scheme and existing works, the average AoI of the proposed scheme is nearly unrelated to CBR while the performance of the other two schemes degrades dramatically with an increasing CBR. When the CBR is larger than $70\%$, the average AoI of the proposed scheme is only $10\%$ of the AoI by the the other two schemes. Meanwhile, we show that the performance of our scheme is near-optimal, validated by comparing the simulations to the theoretical optimum. On the other hand, with the simulation results in the freeway scenario, the proposed scheme still exhibits better performance than the other schemes. Meanwhile, more practical problems are analyzed, which further explains the feasibility of the proposed scheme.
    \end{itemize}
    
    The rest of this paper is organized as follows. SPS scheme in the standard is introduced in Section~\ref{sec_sps_standard} and the related works are presented in Section~\ref{sec_related_works}. The description of the proposed scheme is provided in Section~\ref{sec_caao} and the performance is calculated and analyzed in Section~\ref{sec_performance_analysis}. In Section~\ref{sec_sr}, simulation results are shown and analyzed. Conclusions are given in Section~\ref{sec_conc}.
    
    \section{SPS Scheme in the Standard}
    \label{sec_sps_standard}
    
\begin{figure}[!t]
        \centerline{\includegraphics[width = 0.85\columnwidth ]{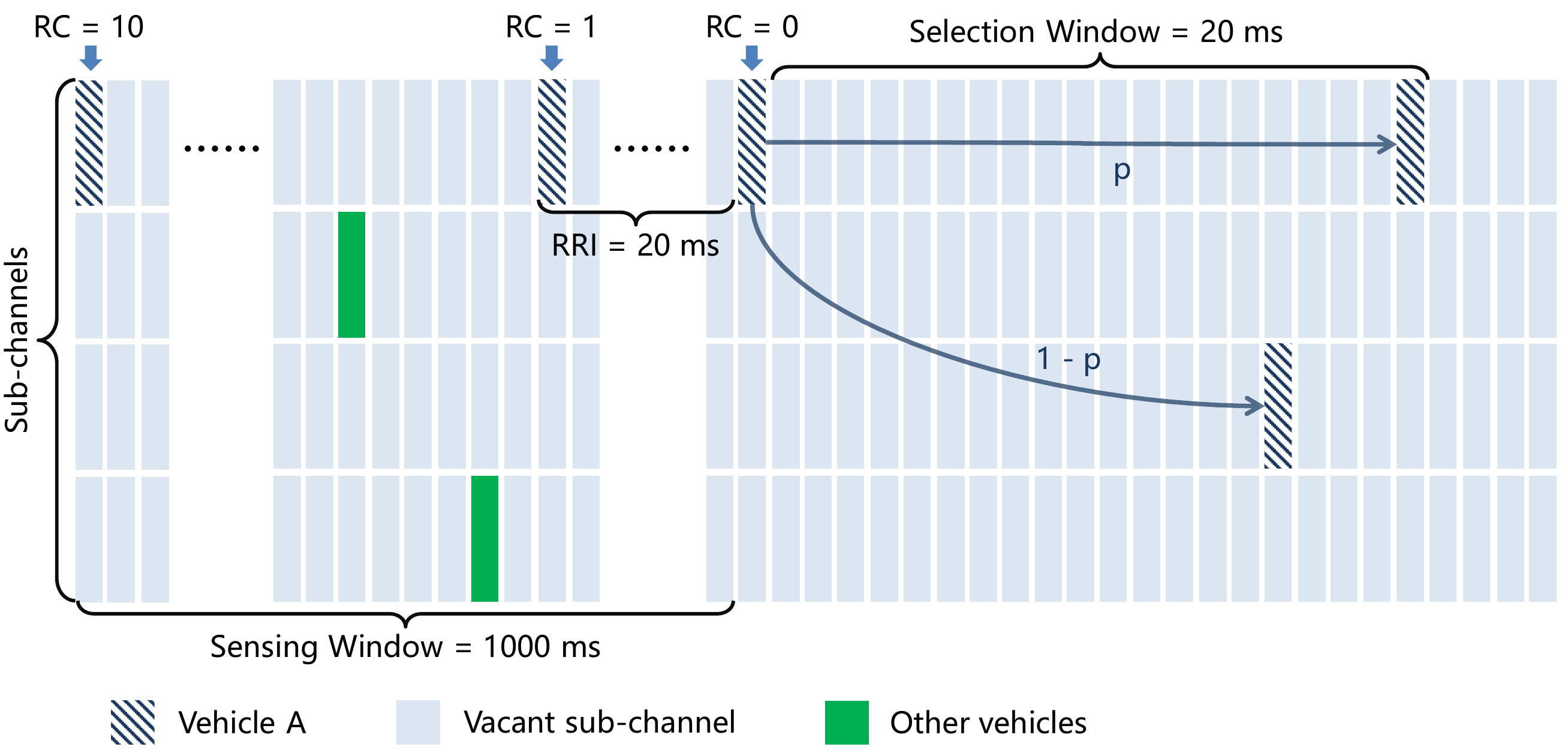}}
        \caption{The process of SPS scheme in 3GPP standard, where the RRI and selection window are $20$ ms.}
        \label{fig:sps_processing}
    \end{figure}

    In the 3GPP C-V2X standard \cite{3gpp:36.213}, the whole bandwidth in a sub-frame is divided into multiple sub-channels, which are composed of a certain number of resource blocks (RB) specified in \cite{3gpp:36.331}. As shown in Fig. \ref{fig:sps_processing}, each vehicle can occupy one or several consecutive sub-channels in frequency domain for transmission according to the data size. 
    
    Based on the structure, the SPS scheme in the standard provides vehicles a distributed method to select resources and broadcast status information. The SPS scheme consists of three steps, i.e., sensing, selection and re-selection. During the communication process, vehicles sense the sub-channel and time slot occupancy in the network. Upon sending a message, a vehicle starts to predict the occupancy in the future resource selection window based on the channel occupancy in the previous $1000$ ms. To be specific, the selection window is located in the future $100$ ms and its size can range from $16$ to $100$, which depends on the implementation. The sub-channels in the selection window will not be considered as candidate resources if meeting the following two conditions. First, if the vehicle has transmitted messages in the last one second, the sub-channels in the corresponding sub-frames in the selection window should be excluded. It is because the occupancy in these sub-frames is unknown due to the HD effect. Second, if the sidelink control information (SCI) was decoded in the last one second and the reference signal received power (RSRP) of the related sub-channels was higher than the threshold, the corresponding sub-channels in the selection window should be excluded as well. After the exclusion, if the proportion of the remainder candidate resources is not lower than $20 \%$ of all resources in the selection window, the vehicle will sort the resources by average received signal strength indication (RSSI) and randomly selects one for transmission among the resources with the lowest $20\%$ RSSI. Otherwise, the threshold will be increased by $3$ dB and the exclusion process will be operated again. 
    
    After the resource is selected, the re-selection counter will be assigned to an initial value, and the vehicle will periodically occupy the corresponding selected resource for transmission. Each time the message is sent, the re-selection counter reduces by one until returning to zero; this is when there will be some probability $1 - p$ for which the vehicle re-selects the resource, which is equivalent to re-executing the selection process of the scheme. An example of the process of the SPS scheme is shown in Fig. \ref{fig:sps_processing}.
    
    \section{Related Works}
    \label{sec_related_works}
    Optimizing AoI in wireless networks has attracted considerable attentions. Ref. \cite{kadota18,hsu18} have considered the wireless broadcast networks wherein scheduling decisions are centralized, and adopted the Whittle's index approach. The wireless multi-access scenario, and hence decentralized scheduling decisions are mode, is considered in \cite{jiang18_iot,jiang18_itc,yates17,kosta18,ali19}. Building on existing MAC-layer protocol, e.g., CSMA or ALOHA, the access probability of backoff window size is optimized in \cite{yates17,kosta18,ali19}, on account of the fact that nodes may have different channel conditions, service rates and packet arrival rates. Ref. \cite{jiang18_itc} adopts the Whittle's index approach and associates the access probability with the index. In \cite{jiang17_iotj}, a round-robin scheduling policy is shown to be asymptotically optimal when the number of nodes is large. In addition, Ref. \cite{talak18,talak_mobihoc} show that a stationary policy actually achieves order-optimal performance in general network topology. However, AoI optimizations in C-V2X are seldom considered in existing works, whereas the MAC layer of C-V2X requires unique treatment.
    
    According to the SPS scheme, each vehicle predicts the future sub-channel and time-slot occupancy in the whole network by sensing historic channels. Similar with most decentralized radio access schemes such as Carrier-Sense Multiple-Access (CSMA), SPS suffers from collisions when the network is congested. Therefore, most works on SPS have been focused on how to reduce the packet collisions. Moreover, MAC layer failures of mode 4 also include other types, e.g., errors due to HD, which are detailed in \cite{gonzalez2019analytical} and \cite{molina2017system}. Based on the analysis, it is pointed out that four types of transmission errors exist in C-V2X mode 4, i.e., HD errors, packet collisions, link-budget deficiency and channel fading caused failures. Among these, the former two are MAC errors which are related to SPS and thus are the focus of this paper.
    
    Since the SPS scheme has been specified in 3GPP Release 14 \cite{3gpp:36.213}, many studies have analyzed its performance. For example, Ref. \cite{nabil2018performance}, \cite{bazzi2018study}, \cite{molina2018configuration} and \cite{toghi2018multiple} analyze the performance of the SPS scheme under various parameters.
    It is pointed out in \cite{nabil2018performance} that the RRI significantly affects the packet data rate, while the parameter of resource keep probability has little effect on the performance in highway scenarios with dense vehicles. 
    Ref. \cite{bazzi2018study} and \cite{molina2018configuration} analyzes more parameters, and indicates that the smaller the size of the sensing window is, the more accurate the prediction on the future channel occupancy will be, but the size of the sensing window cannot be less than the minimum RRI. Meanwhile, it is claimed in \cite{bazzi2018study} that in a V2X network with low congestion, the optional resources need to be maintained at $20\%$ of the total resources, and for a network with very high congestion, the ratio can be adjusted to $10\%$. 
    In \cite{toghi2018multiple}, a detailed assessment of C-V2X technology under high-density vehicular scenarios is presented.
    
    Although the SPS scheme has been specified by 3GPP, it is far from perfect and extensive work has been dedicated to reduce its MAC error rate, which is currently unsatisfactory and cannot meet the reliability requirements for V2X transmissions.
    An optimization scheme is proposed by \cite{molina2017lte} based on the characteristics of the periodicity of safety messages and the regularity of the size packets. It is claimed that the safety information is often composed of a small number of long packets and a large number of short packets, and it is proposed to perform SPS only for short packets and dynamic scheduling for long packets. 
    Different from calculating the average value of the energy of the corresponding sub-channels in sensing window in standard, Ref.\cite{abanto2018enhanced} highlights the proportion of the latest channel occupancy state though exponentially weighting the energy of the corresponding sub-channels in the time domain and achieves a better performance.
    From the perspective of transmitting power, it is found in \cite{kang2018sensing} that higher transmitting power responds to a larger communication range while more easily leads to channel interference, especially under the high channel busy ratio. With transmitting power decreasing, the reception correct rate increases at the cost of communication range. Therefore, Ref.\cite{kang2018sensing} proposes to estimate the channel busy ratio according to the strength of the signaling energy and uses the estimated channel busy ratio to calculate corresponding transmitting power, which improve the performance compared with the standard.
    Based on \cite{kang2018sensing}, Ref.\cite{haider2019adaptive} provides another algorithm whose performance is better than \cite{kang2018sensing}.
    In order to improve the reliability of SPS, the concept of collaboration is introduced in \cite{bonjorn2018enhanced}, whereby vehicles receive the values of the re-selection counter of other vehicles when sensing the channel. Based on these values, each vehicle can adjust its own re-selection counter to minimize the packet collisions.    
    On the basis of \cite{bonjorn2018enhanced}, the SPS scheme with collaboration is further improved in \cite{jeon2018reducing}, wherein the value of the re-selection counter is replaced by the resource location of next re-selection of each vehicle, so that each vehicle can improve the reliability by avoiding the resources that will be occupied by other vehicles according to the collaboration messages. However, only packet collision is considered in both Ref.\cite{bonjorn2018enhanced} and \cite{jeon2018reducing}, where the errors caused by HD effect are neglected.
    Ref.\cite{molina2019geo} establishes a geo-based scheduling to mitigate hidden-terminal problem and reduce the probability of the packet collisions, where vehicles occupy fixed sub-channels according to the relative locations. Nevertheless, stable relative locations may lead to persistent deafness to the vehicles transmitting in the same sub-frame due to HD effect.

    \section{Proposed Collision Avoidance Scheme}
    \label{sec_caao}
    
    Different from SPS-based distributed resource allocation schemes, our proposed scheme does not rely on random re-selection to avoid resource collisions. Instead, it adopts piggyback feedback to avoid collision error. Thereby, each vehicle selects and keeps a fixed resource broadcasting data periodically until a message for collaboration is received. The message is piggybacked with outgoing data from a vehicle, which is transmitted for collision avoidance. Once a vehicle receives the message for collaboration, it will enter re-selection process, which will be described in detail in \ref{subsec:sel}. Therefore, the proposed scheme in this paper is named CAPS for Collision Avoidance based Persistent Scheduling.
    Since fixed occupancy may result in persistent deafness between two vehicles transmitting in the same sub-frame due to HD effect, sub-frame offset method is adopted in CAPS, where vehicles periodically offset the mapping locations according to a certain rule. The offset method is stated in Section \ref{subsec:subf_offset}. In addition, since each vehicle tends to persistently keep a sub-channel for transmission in CAPS, the number of vehicles should be less than the total number of sub-channels in a period, which can be guaranteed in our scheme.
    
\begin{figure}[t]
        \centerline{\includegraphics[width = 0.85\columnwidth ]{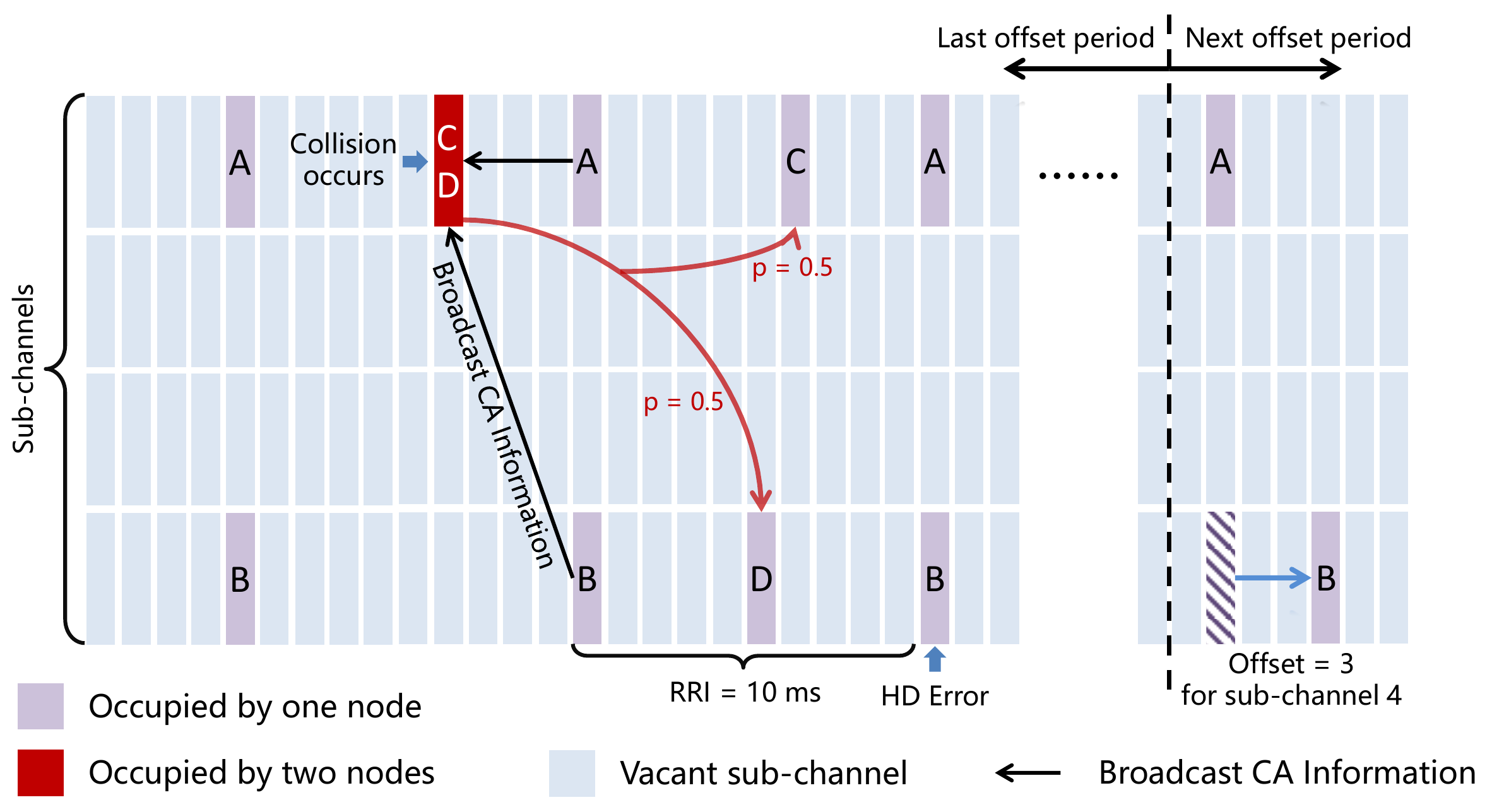}}
        \caption{An example of the process of CAPS scheme. }
        \label{fig:ca}
    \end{figure}

    An example of the process of CAPS scheme is shown in Fig. \ref{fig:ca}. Specifically, the figure shows the process of collision avoidance and sub-frame offset. In this example, there are four vehicles transmitting messages in their own sub-channel, where vehicle C and D are transmitting in the same sub-channel noted by the red area. For vehicle A and B, they can sense the occupancy of the red sub-channel but cannot decode it. Then the red area is suspected of collision and vehicle A and B will broadcast the collaboration message together with the payload when transmitting. If vehicle C and D hear the collaboration message, they will re-select a new sub-channel with a certain probability. In this example, vehicle C keeps the same sub-channel and vehicle D re-selects a new sub-channel.
    
    In the remainder of this section, we will explain CAPS scheme consisting of four steps in detail, i.e., sensing, collaboration, sub-frame offset and selection/re-selection.
    
    \subsection{Sensing}\label{subsec:sensing}
    
    The sensing process is designed to sense the occupancy of the sub-channels during a certain time of the past. Based on that, each vehicle can predict the occupancy of sub-channels in the future when selecting or re-selecting a new resource for transmission. Meanwhile, each vehicle could judge whether a collision occurs in a sub-channel through sensing process. In CAPS scheme, the sensing window length is designed to be two times of the RRI of each vehicle, e.g., a length of $40$ ms for an RRI of $20$ ms, which is different from the sensing window length of one second designed in the standard SPS scheme. One reason is that the longer the sensing window is, the less prominent the latest resource occupancy is; the re-selection, therefore, would result in an inability to make accurate predictions on future sub-channel occupancy. The other reason is related to sub-frame offset adopted in CAPS scheme, which is specified in detail in Section \ref{subsec:subf_offset}. With the sensing window length shortening, there is a certain risk that the occupancy of a packet with a longer RRI cannot be sensed. Benefiting from the collaboration mechanism in CAPS that is stated in Section \ref{subsec:collaboration}, the packet collision caused by this reason can be quickly solved, so the impact on the incomplete sensing of sub-channel occupancy can be neglected.
    
    At each time slot, the vehicle judges the sub-channel occupancy in the sensing window in terms of the signal strength of the sub-channels. Because of the fixed occupancy of resources in CAPS scheme, once a sub-channel is sensed to be occupied, the corresponding resource is considered to be occupied persistently.
    
    \subsection{Collaboration}\label{subsec:collaboration}
    
    For distributed resource allocation, sensing-based prediction is not always accurate due to the uncertainty of the channel selection in the future, which may result in packet collision. Therefore, sensing-based only scheme, such as the standard SPS scheme, cannot lead to good performance. Meanwhile, because of the semi-persistent scheduling scheme, the collision persists until next re-selection without being known by the transmitters. To enhance the performance, it is necessary for each vehicle to obtain more information about the channel status.
    
    Therefore, the idea of collaboration is introduced in our scheme and the end result in terms of performance turns out much better than conventional schemes based on simulation results in Section \ref{sec_sr}. To achieve collaboration among vehicles, the messages for collaboration are transmitted for collision avoidance via piggyback, which can help reducing the number of packet collisions in the network.
    \begin{defi}[Suspected of Collision]
    \label{defi_col_sus}
    If the energy of the sub-channel is higher than the threshold in the sensing window, but its content cannot be decoded correctly, then the sub-channel will be considered suspected of collision. 
    \end{defi}
    
    If a vehicle has observed this and needs to broadcast safety messages in the current sub-frame, the message for collision avoidance will be transmitted together with the payload, which indicates the time and frequency location of the sub-channels that are suspected of collisions. If the vehicle transmitting in the corresponding sub-channel receives the message, it will enter re-selection process. In addition, in order to avoid repeatedly handling the messages for the same sub-channel from multiple vehicles, the message for one sub-channel is designed to be handled only once. 
    \begin{defi}[Assistance Range]
    Assistance range is defined as a past period of time before the collaboration message transmitted and only the sub-channels in the assistance range can be helped to avoid collisions. In CAPS, the assistance range is designed not to exceed the respective RRI of each vehicle.
    \end{defi}
    
    \subsubsection{Signaling Overhead}
    It is noted that with the introduction of messages for collaboration, the signaling overhead is bound to increase with the number of vehicles and packet collisions. In order to guarantee data throughput while improving reception accuracy, it is necessary to consider the signaling overhead of the scheme and limit the overhead to an acceptable range. In our scheme, a message for collaboration indicates the time and frequency locations of the sub-channels by the index of sub-frames and sub-channels. Therefore, the bit length $l_\text{ca}$ of the location field for one sub-channel depends on the assistance range $N_\text{subf}$ and the number of sub-channels $N_\text{subCH}$ in one sub-frame and can be calculated by
    \begin{equation}\label{equ:size_ca}
    l_\text{ca} = \left \lceil \log_{2} N_\text{subf} \right \rceil + \left \lceil \log_{2} N_\text{subCH} \right \rceil.
    \end{equation}
    
    If the RRI of a vehicle is $50$ ms and there are $50$ sub-channels in one sub-frame, which means the bit length for one sub-channel is equal to $12$ bits. To limit the overhead to an acceptable range, each vehicle is allowed to broadcast three sub-channels suspected of collision at most in CAPS. Therefore, the overhead of the collaboration message equals to $36$ bits at most.
    
    In addition, as a side benefit, since packet collisions are notified by the receiving vehicles, CAPS scheme can also mitigate the hidden terminal problem.
    
    \subsection{Sub-frame Offset}\label{subsec:subf_offset}
    
\begin{figure}[t]
        \centerline{\includegraphics[width = 0.85\columnwidth ]{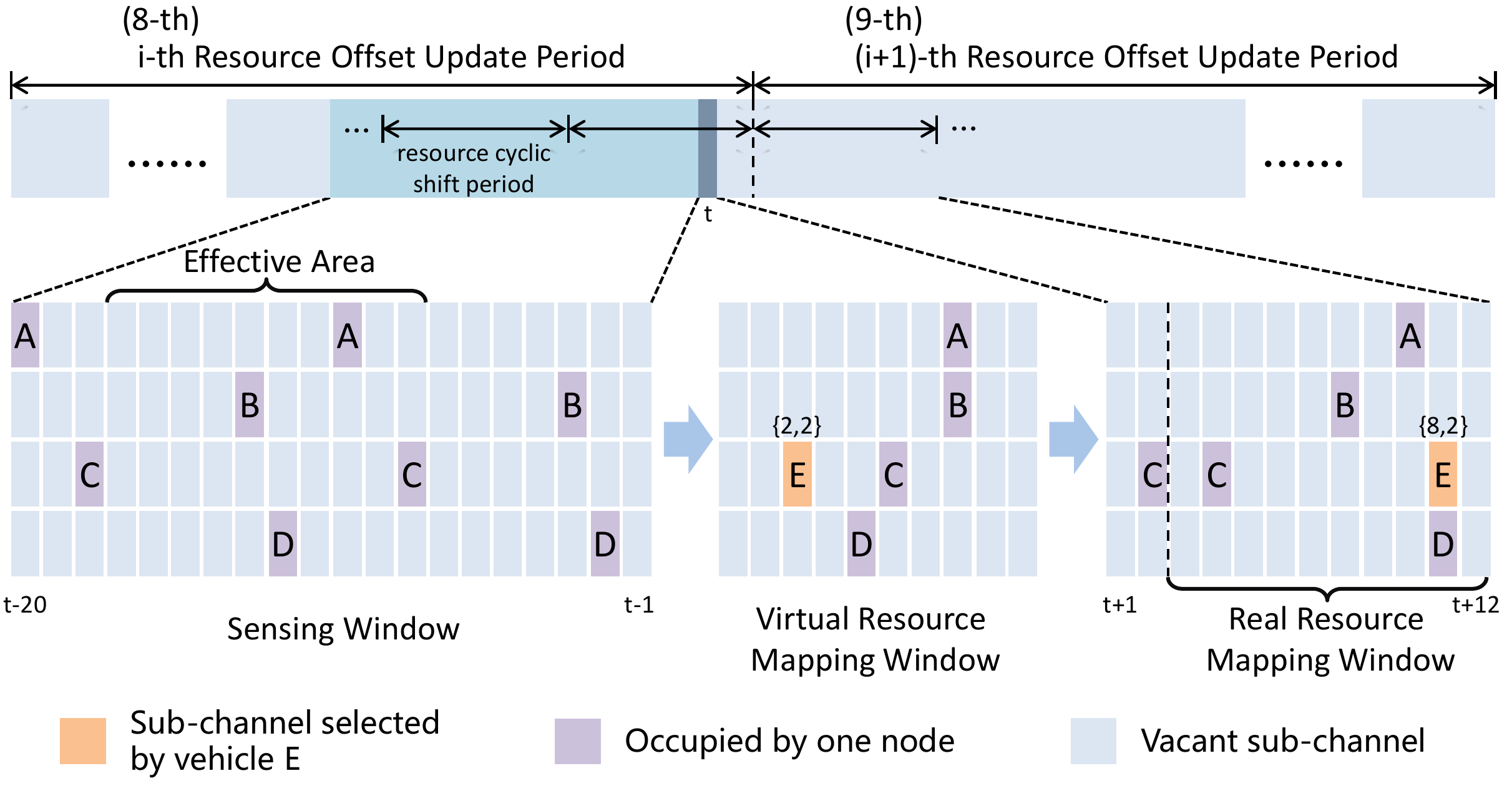}}
        \caption{An example of selection process with sub-frame offset method, where $i=8, \alpha_{\text{RRI}}=10, T_{\text{ost}}=10 \text{ ms and } T_{\text{upd}}=1000 \text{ ms}$. The length of sensing window equals to two times of RRI, i.e. $20$ ms, and the length of effective area equals $10$ ms. In the current sub-frame $t$, vehicle E would like to select a new resource for transmission. First, the locations of the resources in the effective area are offset to their original locations in the virtual resource mapping window. When vehicle E selects the sub-channel located in the $3$-rd sub-frame and the $3$-rd sub-channel \{$2,2$\} in the virtual window, the corresponding sub-channel in the real resource mapping window will be mapped to the location of the $9$-th sub-frame and the $3$-rd sub-channel \{$8,2$\} in the $9$-th resource offset update period. Since the process stretches across two adjacent resource offset update periods, the locations of the other resources are offset as well.}
        \label{fig:sub-frame offset}
    \end{figure}

    With the collaboration mechanism, vehicles always occupy the fixed resources for transmission until receiving the message for collaboration. When there is no packet collision and no collaboration message is transmitted, the vehicles transmitting in the same sub-frame would never hear each other due to HD effect. In practice, the persistent deaf phenomenon can lead to a high risk of safety. To mitigate the problem, a periodic sub-frame offset method is introduced. Using the method, the mapping of different sub-channels in the same sub-frame will be staggered in the time domain according to a certain rule. To implement the offset method, virtual resource mapping window and real resource mapping window are designed in CAPS.
    \begin{defi}[Virtual Resource Mapping Window and Real Resource Mapping Window]
    Virtual resource mapping window is defined as a logical window of each vehicle where the resource occupancy of the surrounding vehicles is always fixed unless collision occurs, while the real resource mapping window reflects the practical occupancy of the resources after sub-frame offset. In the other word, the resource occupancy in the real window is periodically staggered in the time domain from the virtual window. The time spans of the virtual and real resource mapping windows are equal to the respective RRI of each vehicle. An example of the two windows is shown in Fig. \ref{fig:sub-frame offset}.
    \end{defi}
    
    Since the direct sub-frame and frame number are broadcasted in Physical Sidelink Broadcast Channel (PSBCH) as specified in the standard \cite{3gpp:36.331}, the time domain information can be adopted to offset sub-frame.
    
    In CAPS, the time domain is divided by resource cyclic shift period $T_\text{ost}$ and resource offset update period $T_\text{upd}$ respectively as shown in Fig. \ref{fig:sub-frame offset}.
    \begin{defi}[Resource Cyclic Shift Period and Resource Offset Update Period]
    Resource cyclic shift period and resource offset update period are system parameters. Resource cyclic shift period is defined as a period where the occupancy location is circularly offset in the time domain. Resource offset update period reflects the offset degree of the resources. In each resource offset update period, the resources in the same frequency domain circularly offset the same number of sub-frames in the corresponding resource cyclic shift period. In addition, $T_\text{upd} = nT_\text{ost} \geq \alpha_{\text{RRI}} = mT_\text{ost}$, where $n, m \in \mathbb{Z}^{+}$ and $\alpha_{\text{RRI}}$ indicates the resource reservation interval.
    \end{defi}
    
    To be specific, assuming that a sub-channel in the virtual resource mapping window is located in \{$n_\text{subf}$, $n_\text{subCH}$\} in the time and frequency domain, if the corresponding sub-channel in real mapping window is located in the $i$-th resource offset update period, the sub-frame offset $O_\text{subf}$ can be calculated by $O_\text{subf} = n_\text{subCH}  (i - 1)$, where $n_\text{subCH} \in \{0,...,N_\text{subCH} - 1\}$, $n_\text{subf} \in \{0,...,N_\text{subf} - 1\}$, $N_\text{subCH}$ and $N_\text{subf}$ respectively indicates the number of sub-channels in one sub-frame and the number of sub-frames in a virtual resource mapping window, and $i$ starts from one. Therefore, the sub-fame of the corresponding sub-channel in the real resource mapping window after offset could be expressed by $\left (n_\text{subf} + O_\text{subf}\right ) \Mod{T_\text{ost}} + \left \lfloor \frac {n_\text{subf}}{T_\text{ost}} \right \rfloor  T_\text{ost}$ and the frequency domain of the sub-channel keeps the same.
    \begin{prop}
    Based on the sub-frame offset method, each sub-channel in the virtual mapping window can correspond a one and only sub-channel in the real mapping window and vice versa.
    \end{prop}
    
    \subsection{Selection/Re-selection}\label{subsec:sel}
    
    When a vehicle would like to select a new resource for transmission at sub-frame $t$, it will first find an effective area in the sensing window.
    \begin{defi}[Effective Area]
    Since the locations of the resources have been offset in the practical channel, it is necessary to calculate the original locations of the resources before the selection process. Therefore, effective area is designed and only the sub-channels in the area can be offset to their original locations. The effective area consists of an integer number of the latest resource cyclic shift periods before sub-frame $t$ as shown in Fig. \ref{fig:sub-frame offset}, whose length is equal to the RRI of the vehicle. If the sensing window stretches across two adjacent resource offset update periods, the effective area consists of the last one or several resource cyclic shift periods in the last resource offset update period.
    \end{defi}
    
    After determining the effective area, the sub-channels in it can be mapped into virtual resource mapping window based on Section \ref{subsec:subf_offset}. In the virtual mapping window, the sub-channels whose energies are under the threshold are considered to be available to select. Meanwhile, the sub-channels that cannot be detected due to HD effect should be excluded. If the number of available sub-channels is zero after filtering, the threshold will increase to guarantee that at least one sub-channel is available. Then the vehicle will randomly select a sub-channel among the set of the available sub-channels and map the selected sub-channel to the two real resource mapping windows after the effective area as shown in Fig. \ref{fig:sub-frame offset}, where the part before the current sub-frame $t$ does not need to be considered and so omitted.
    
    After selecting or re-selecting a new sub-channel, the vehicle will persistently keep the sub-channel and periodically transmitting data until receiving a message for collaboration. When the vehicle enter the next resource offset update period, it will offset the sub-channel in the time domain according to Section \ref{subsec:subf_offset}.
    
    For resource re-selection, note that entering the re-selection process does not absolutely equal to re-select a new sub-channel. In our scheme, considering that there are at least two packets in a collision sub-channel, the probability of re-selection for collision avoidance is set to be $0.5$.
    
    The complete procedure of CAPS for each vehicle is shown in Algorithm \ref{algorithm} in detail. In line 1, $n_\text{subCH}$ indicates the index of the sub-channel and $n_\text{csubf}$ indicates the current sub-frame. In line 2, $T_\text{simu}$ indicates the duration of the simulation. In line 3, it is judged whether the data is transmitted in the current sub-frame and the sub-frame for transmission is expressed by $n_\text{subf\_data}$. In line 8, the number of sub-channels in a sub-frame is shown by $N_\text{subCH}$. In line 11, $E_\text{subCH}$ indicates the energy of current sub-channel and $E_\text{Thre}$ indicates the threshold for collision detection. In line 14, it is judged whether the message for collaboration exists in the packet and the location indicated in the message points to the own sub-channel.
    
    \begin{algorithm}[!t]
    \SetAlgoLined
        Initialization: $n_\text{subCH} = 0$; $n_\text{csubf} = 0$\;
        \While{$n_\text{csubf} \leq T_\text{simu}$}{
            \eIf{$n_\text{subf\_data} == n_\text{csubf}$}{
                Transmit data in corresponding sub-channel\;
                Transmit message for collaboration in corresponding sub-channel; \text{ }\text{ }$\triangleright\text{ }Section\text{ }\ref{subsec:collaboration}$\\  
                Calculate the location of the next sub-channel for transmission; \text{ }\text{ }\text{ }\text{ }\text{ }$\triangleright\text{ }Section\text{ }\ref{subsec:sel}$\\  
                }{
                \While{$n_\text{subCH} < N_\text{subCH}$}{
                    Sense the energy of the sub-channel; \text{ }\text{ }\text{ }\text{ }\text{ }\text{ }\text{ }\text{ }\text{ }\text{ }\text{ }\text{ }\text{ }\text{ }\text{ }\text{ }\text{ }\text{ }\text{ }\text{ }\text{ }\text{ }\text{ }\text{ }\text{ }\text{ }\text{ }\text{ }\text{ }\text{ }\text{ }\text{ }$\triangleright\text{ }Section\text{ }\ref{subsec:sensing}$\\ 
                    Receive and decode the sub-channel\;
                    \uIf{$E_\text{subCH} > E_\text{Thre}$ but CRC check is False}{
                        The sub-channel is suspected of collision and cached; \text{ }\text{ }\text{ }\text{ }\text{ }\text{ }\text{ }$\triangleright\text{ }Section\text{ }\ref{subsec:collaboration}$\\  
                    }
                    \ElseIf{$E_\text{subCH} > E_\text{Thre}$ and CRC check is True}{
                        \If{message for collaboration exists and is valid for myself}{ 
                            Enter the process of re-selection\;
                            Map effective area to virtual window; \text{ }\text{ }\text{ }\text{ }\text{ }\text{ }\text{ }\text{ }\text{ }\text{ }\text{ }\text{ }\text{ }\text{ }\text{ }\text{ }\text{ }\text{ }\text{ }\text{ }\text{ }$\triangleright\text{ }Section\text{ }\ref{subsec:sel}$\\  
                            Select a new sub-channel; \text{ }\text{ }\text{ }\text{ }\text{ }\text{ }\text{ }\text{ }\text{ }\text{ }\text{ }\text{ }\text{ }\text{ }\text{ }\text{ }\text{ }\text{ }\text{ }\text{ }\text{ }\text{ }\text{ }\text{ }\text{ }\text{ }\text{ }\text{ }\text{ }\text{ }\text{ }\text{ }\text{ }\text{ }\text{ }$\triangleright\text{ }Section\text{ }\ref{subsec:sel}$\\ 
                            Map the new sub-channel from virtual window to real window; \text{ }\text{ }\text{ }\text{ }\text{ }\text{ }\text{ }\text{ }\text{ }\text{ }\text{ }\text{ }\text{ }\text{ }\text{ }\text{ }\text{ }\text{ }\text{ }\text{ }\text{ }\text{ }\text{ }\text{ }\text{ }\text{ }\text{ }\text{ }\text{ }\text{ }\text{ }\text{ }\text{ }\text{ }\text{ }\text{ }\text{ }\text{ }\text{ }\text{ }\text{ }\text{ }\text{ }\text{ }\text{ }\text{ }\text{ }\text{ }\text{ }\text{ }\text{ }\text{ }\text{ }\text{ }\text{ }\text{ }\text{ }\text{ }\text{ }\text{ }\text{ }\text{ }\text{ }\text{ }\text{ }$\triangleright\text{ }Section\text{ }\ref{subsec:sel}$\\  
                        }
                    }
                    $n_\text{subCH} = n_\text{subCH} + 1$\;
                }
                $n_\text{subCH} = 0$\;
            }
            $n_\text{csubf} = n_\text{csubf} + 1$\;
        }
        \caption{CAPS Scheme}\label{algorithm}
    \end{algorithm}

    \section{Performance Analysis}
    \label{sec_performance_analysis}
    
    In this section, performance of CAPS scheme is analyzed theoretically with closed-form expressions. As mentioned in Section \ref{sec_caao}, CAPS scheme aims to converge to a stable status, where there is no packet collision and each vehicle keeps the relatively fixed sub-channel for periodical transmission during a period. In this section, the performance analysis is based on the converged status of the scheme, which means that collaboration and re-selection process do not need to be considered in the analysis. Therefore, before the performance analysis, the convergence analysis is necessary that is detailedly analyzed in Appendix \ref{sec_proof}.
    
    In order to intuitively evaluate the performance of CAPS scheme, AoI is introduced to be the performance evaluation index, since it can integrally reflect reliability and latency and is suitable for evaluating the performance of periodical transmission. To be specific, the average AoI of reception of each vehicle is considered to evaluate the performance of CAPS scheme. Besides, the calculation of AoI is discrete and the minimum unit of AoI is one millisecond. The performance of CAPS scheme under both static and dynamic vehicular traffic flows is analyzed in detail in what follows.
    
    \subsection{Analysis of Static Vehicular Traffic Flow}\label{subsec:static}
    In the scenario of static vehicular traffic flow, vehicles in one communication range will not leave the range and no new vehicles will enter the range, so the number of the vehicles in the range is constant as well. When the scheme is converged, the average AoI of reception of each vehicle can be expressed by
    \begin{equation}\label{equ:average_AoI_1_main}
        \mathbb{E}_{i} \left[a_{i}\right] = \frac{1}{v} \sum_{i=1}^{v} a_{i},
    \end{equation}
    where $v \leqslant c$. In addition, the variable $v$ and $c$ respectively indicate the total number of vehicles and the number of sub-channels during an RRI, the variable $a_{i}$ indicates average AoI of reception of AoI and $i$ indicates the receiving vehicle $i$. After the calculation and analysis in Appendix \ref{sec_static}, Theorem \ref{theo:static} is obtained.
    \begin{tcolorbox}
    \begin{theorem}[Performance of Static Vehicular Traffic Flow]\label{theo:static}
    In the scenario of static vehicular traffic flow, after CAPS converges, the average AoI of reception of each vehicle is calculated by
    \begin{iarray}\label{equ:E_i_static_main}
        \mathbb{E}_{i} [a_{i}] &=& \frac{(N_\text{subCH}-1)(T+\alpha_{\text{RRI}})}{2(c-1)} + \frac{(c-N_\text{subCH})(\alpha_{\text{RRI}}-1)}{2(c-1)},
    \end{iarray}
    where $N_\text{subCH}$ indicates the number of sub-channels in a sub-frame, variable $T$ indicates the period where relative locations of the sub-channels occupied by vehicles keep the same, $c$ equals to the number of sub-channels during an RRI and $\alpha_{\text{RRI}}$ indicates the resource reservation interval. Meanwhile, the total number of vehicles $v$ should be equal or less than $c$.
    \end{theorem}
    \end{tcolorbox}
    Shown by the analysis results in (\ref{equ:E_i_static_main}), the average AoI in static scenario is irrelevant of the number of vehicles $v$ or CBR, when $v \leqslant c$ is met and CAPS is converged. It is because there is no packet collision after CAPS converges, the only MAC error is caused by HD effect, which is not affected by the number of vehicles $v$ when sub-frame offset method is used.

    \subsection{Analysis of Dynamic Vehicular Traffic Flow}\label{subsec:dynamic}
    In the scenario of dynamic vehicular traffic flow, it is assumed that vehicles in one communication range will arrive in and leave the range in a dynamic rate, but the number of the vehicles in the range is still constant. In the performance analysis of this part, the expectation of the proportion of the number of vehicles arriving and leaving a communication range in one unit of time are respectively equal to $x$ and $y$. For instance, when the number of vehicles in a communication range equals $100$, dynamic rate $x=y=0.7$ indicates that there are $100 \times 0.7 = 70$ new vehicles arriving in and $70$ vehicles leave the range in one second. Similarly, the scenario of static vehicular traffic flow responds to $x=y=0$. In order to simplify the calculation, it is assumed that 
    \begin{itemize}
        \item both $x$ and $y$ are constant and $x=y$,
        \item the minimum calculation window is an RRI,
        \item the scheme can be converged in each RRI,
        \item the packet will not collide if it has collided in the last RRI.
    \end{itemize}
    
    Based on the assumptions mentioned above, analysis in detail is shown in Appendix \ref{sec_dynamic} and the analysis results is given in Theorem \ref{theo:dynamic}.
    \begin{tcolorbox}
    \begin{theorem}[Performance of Dynamic Vehicular Traffic Flow]\label{theo:dynamic}
    In the scenario of dynamic vehicular traffic flow, the average AoI of reception of each vehicle is finally expressed by 
    \begin{iarray}\label{equ:average_AoI_final_main}
        \mathbb{E}_{i} [a_{i}] &=& \frac{(N_\text{subCH}-1)(T+\alpha_{\text{RRI}})}{2(c-1)} + \frac{(c-N_\text{subCH})(\alpha_{\text{RRI}}-1)}{2(c-1)} + \frac{(c-N_{\text{subCH}})\alpha_{\text{RRI}}}{(v_{\text{0}}-1)(c-1)}  n_{\text{col}},
    \end{iarray}
    where $N_\text{subCH}$ indicates the number of sub-channels in a sub-frame, variable $T$ indicates the period where relative locations of the sub-channels occupied by vehicles keep the same, $c$ equals to the number of sub-channels during an RRI, $\alpha_{\text{RRI}}$ indicates the resource reservation interval, $v_{\text{0}}$ indicates the initial number of vehicles and $n_{\text{col}}$ indicates the number of collision packets in an RRI window. Meanwhile, the total number of vehicles $v$ should be equal or less than $c$.
    \end{theorem}
    \end{tcolorbox}
    According to Theorem \ref{theo:dynamic}, only the dynamic part of average AoI is related to the initial number of vehicles $v_{\text{0}}$. Compared with the first two parts in (\ref{equ:average_AoI_final_main}), the dynamic part have little effect on the result of the average AoI. Therefore, the average AoI of reception of each vehicle in the scenario of dynamic vehicular traffic flow have little relation with the number of vehicles.
    
    \subsection{Results of Analysis}\label{subsec:optimal_config}
\begin{figure}
    \centering
    \subfigure[]{
    \begin{minipage}[c]{0.48\linewidth}
    \centering
    \includegraphics[width=7.79cm]{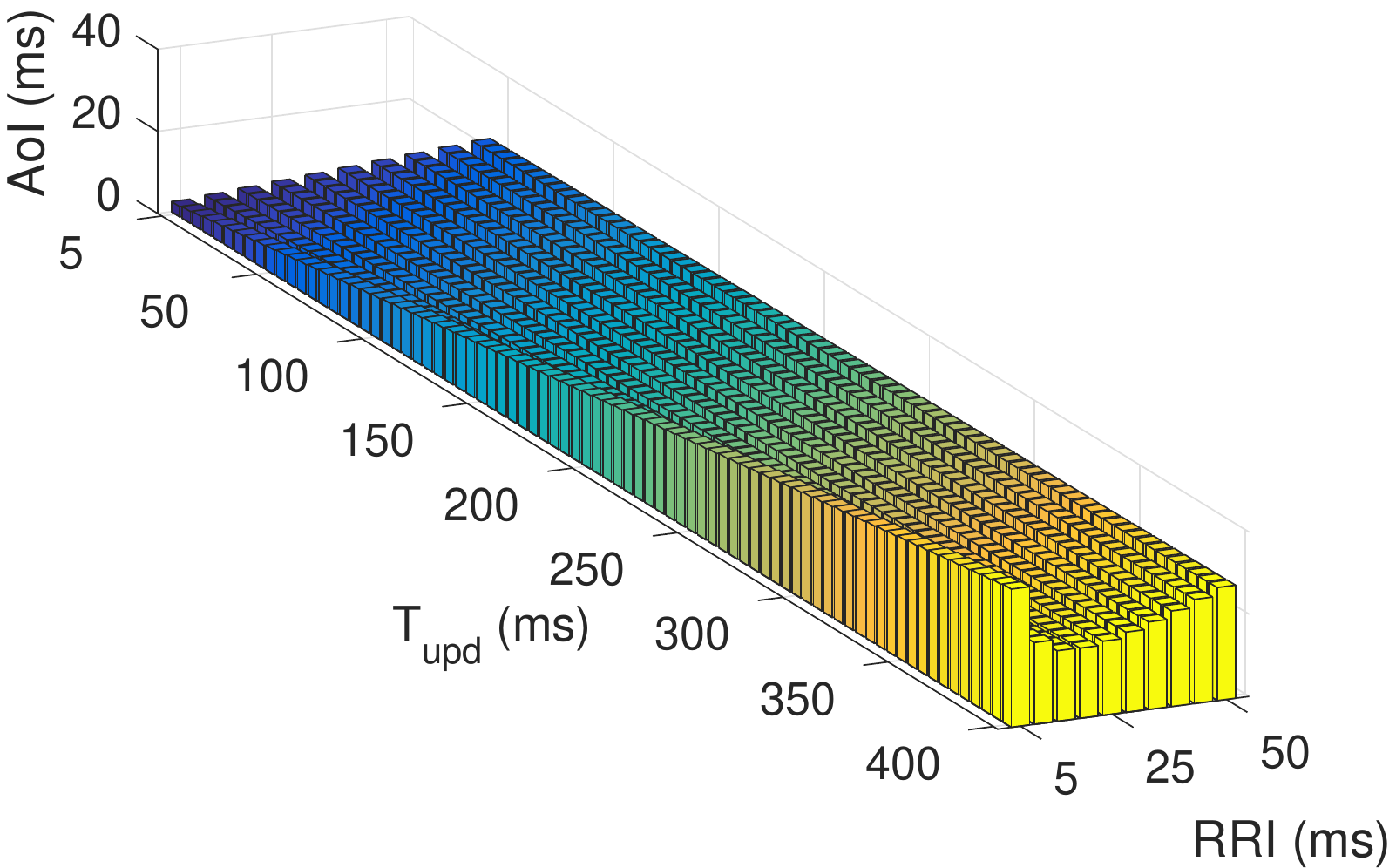}
    \label{fig:aoi_analysis_static}
    \vspace*{-2pt}
    \end{minipage}}
    \subfigure[]{
    \begin{minipage}[c]{0.48\linewidth}
    \centering
    \includegraphics[width=7.9cm]{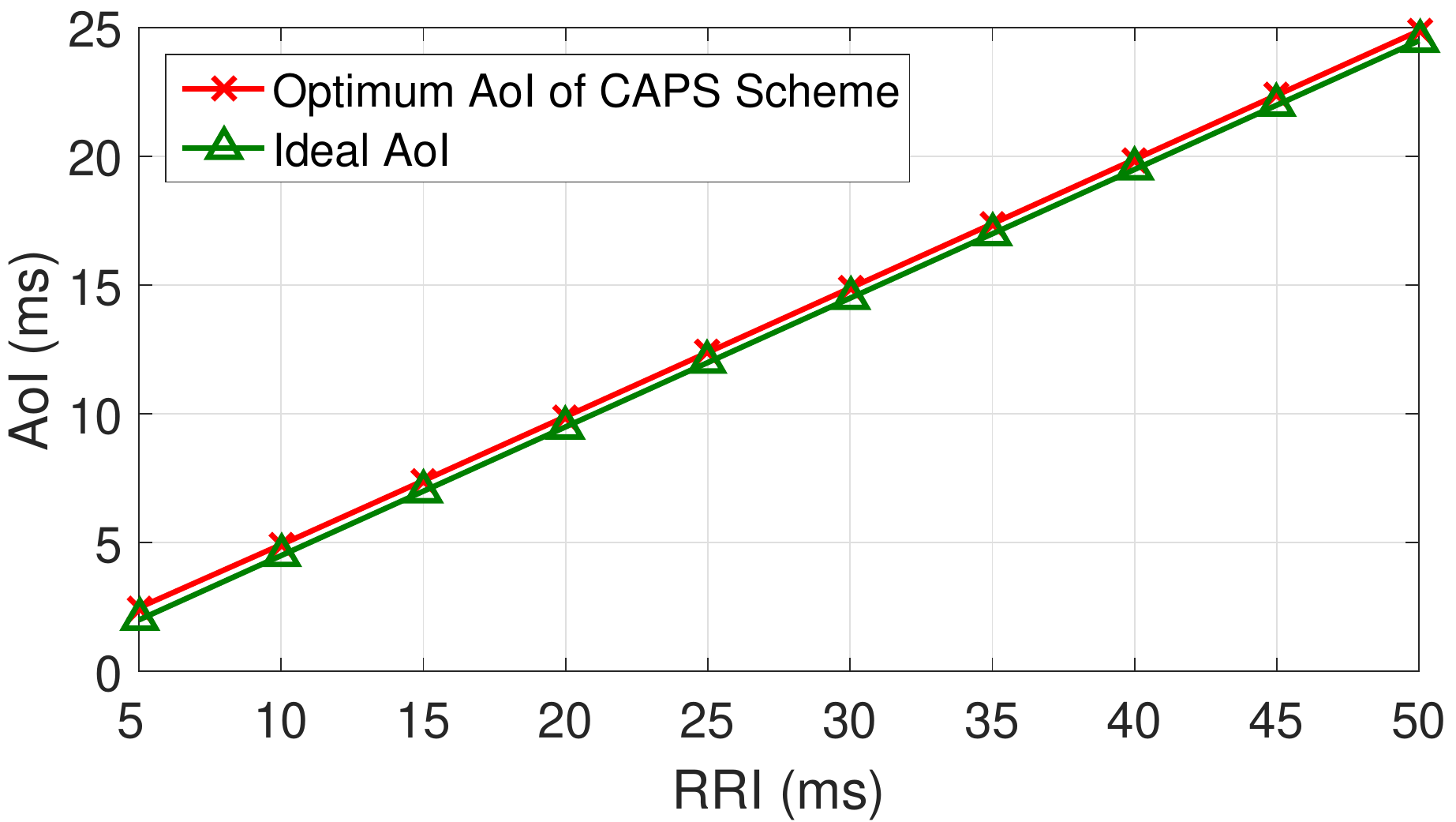}
    \label{fig:aoi_comparison_ideal_and_optimum}
    \vspace*{-10pt}
    \end{minipage}}
    \vspace*{-5pt}
    \caption{(a) Average AoI versus $T_\text{upd}$ and RRI in the scenario of static vehicular traffic flow. (b) Comparison between Optimum AoI of CAPS Scheme and Ideal AoI with a Certain RRI.}
    \end{figure}

    As mentioned above, $T$ indicates the period where relative locations of the sub-channels occupied by vehicles keep the same. Since the sub-frame offset method is adopted in CAPS, the mappings of the resources are relatively fixed in each resource offset update period, so that $T$ equals to a resource offset update period, i.e. $T = T_\text{upd}$. Therefore, $T$ can be replaced by $T_\text{upd}$. When the number of vehicles is less than the total number of sub-channels in a period of transmission and $N_{\text{subCH}}$ is set to $4$, the average AoI can be calculated and the result versus $T_\text{upd}$ and RRI is shown in Fig. \ref{fig:aoi_analysis_static}.
    
    From Fig. \ref{fig:aoi_analysis_static}, we find that with the same RRI, the shorter $T_\text{upd}$ corresponds to the lower AoI, where $T_\text{upd} \geqslant \alpha_{\text{RRI}}$. When $T_\text{upd} = \alpha_{\text{RRI}}$, AoI is minimum and is near the ideal AoI, as shown in Fig. \ref{fig:aoi_comparison_ideal_and_optimum}. The ideal AoI here reflects the ideal scenario where any vehicle can receive the messages from other vehicles. The ideal average AoI can be expressed by $\mathbb{E}_{i} [a_{i}]_{\text{idl}} = \frac{\alpha_{\text{RRI}}-1}{2}$.

    From the other perspective, when $T_\text{upd}$ is set to a constant, there is an RRI corresponding a minimum AoI. Moreover, since $T_\text{upd}$ is always a global parameter and $\alpha_{\text{RRI}}$ is a local parameter in practical scenarios, based on the performance analysis, the optimal RRI can be adopted when $T_\text{upd}$ is determined and $v \leqslant c$. In the end of Appendix \ref{sec_static}, the method of calculating optimal RRI is provided. When $T_\text{upd} = 400 \text{ ms}, N_{\text{subCH}} = 4, T_{\text{ost}} = 5 \text{ ms}$, the theoretically optimal RRI equals to $17.59$ ms. Considering that RRI is a integer and the set of the available RRI is discrete, the practically optimal RRI equals to $20$ ms, which means the RRI of $20$ ms is the optimal and available configuration to achieve the minimum average AoI.

	\section{Simulation Results}
    \label{sec_sr}
    In this section, the simulation results are given and the simulation consists of two parts. The first part is about the simulation on MAC layer where only packet collisions and HD effects are considered. In the second part, the simulation is operated in a more practical scenario where the models of vehicle mobility, physical channel and the loss of physical layer process are included as well.
    
    \subsection{Simulations on MAC layer}
    \label{subsec_sim_mac}
    In the simulations on MAC layer, the comparison with relevant schemes including state-of-the-art is provided. The theoretical analysis and simulation results are also compared in this part. In addition, the conclusion of the optimal RRI with a fixed $T_\text{upd}$ is verified and the range of application of CBR is revised.
    
    \subsubsection{Comparison among Different Schemes}
    In the part of performance comparison, we implement CAPS scheme, the standard SPS scheme and SPS/LA scheme from \cite{jeon2018reducing} on MATLAB, and provide the comparisons through Monte-Carlo simulations. The system bandwidth is configured to be $10$ MHz and there are $4$ sub-channels of $12$ RBs in the whole bandwidth. Therefore, the message of 190 bytes needs one sub-channel for transmission. Some related parameters are listed in Table~\ref{Simulation Settings}. In the simulations, all the vehicles broadcast their messages by sharing the $10$ MHz bandwidth. Since only the MAC error is considered in this part, the physical-layer errors due to, e.g., channel fading, are omitted here. In addition, HD influence is fully considered during sensing and collaboration, so the performance of each scheme can be more accurately reflected. For initialization, vehicles randomly select sub-channels to transmit data in one second.
    
    \begin{table} [t] 
    \centering 
    \caption{Simulation Settings for CAPS Scheme} 
    \label{Simulation Settings}
    \footnotesize
    \begin{tabular}{c c|c c}  
    \hline
    \textbf{Parameter}&\textbf{Value}&\textbf{Parameter}&\textbf{Value}\\
    \hline
    Number of Vehicles ($v$) & $2$ to $180$ & Resource Offset Update Period ($T_{\text{upd}}$) & $50/200/1000$ ms \\
    \hline
    RRI ($\alpha_{\text{RRI}}$) & $50$ ms & Traffic Flow of Vehicles & Static \\
    \hline
    Number of Sub-channels ($N_\text{subCH}$) & $4$ & Maximum Number of Collision Assisted & $3$ \\
    \hline
    Duration of Simulation & $100$ s & Re-selection Probability for Collision & $0.5$ \\
    \hline
    Resource Cyclic Shift Period ($T_{\text{ost}}$) & $10$ ms & Message Size & $190$ bytes \\
    \hline
    \end{tabular}  
    \end{table} 
    
\begin{figure}[t]
        \centerline{\includegraphics[width = 0.82\columnwidth]{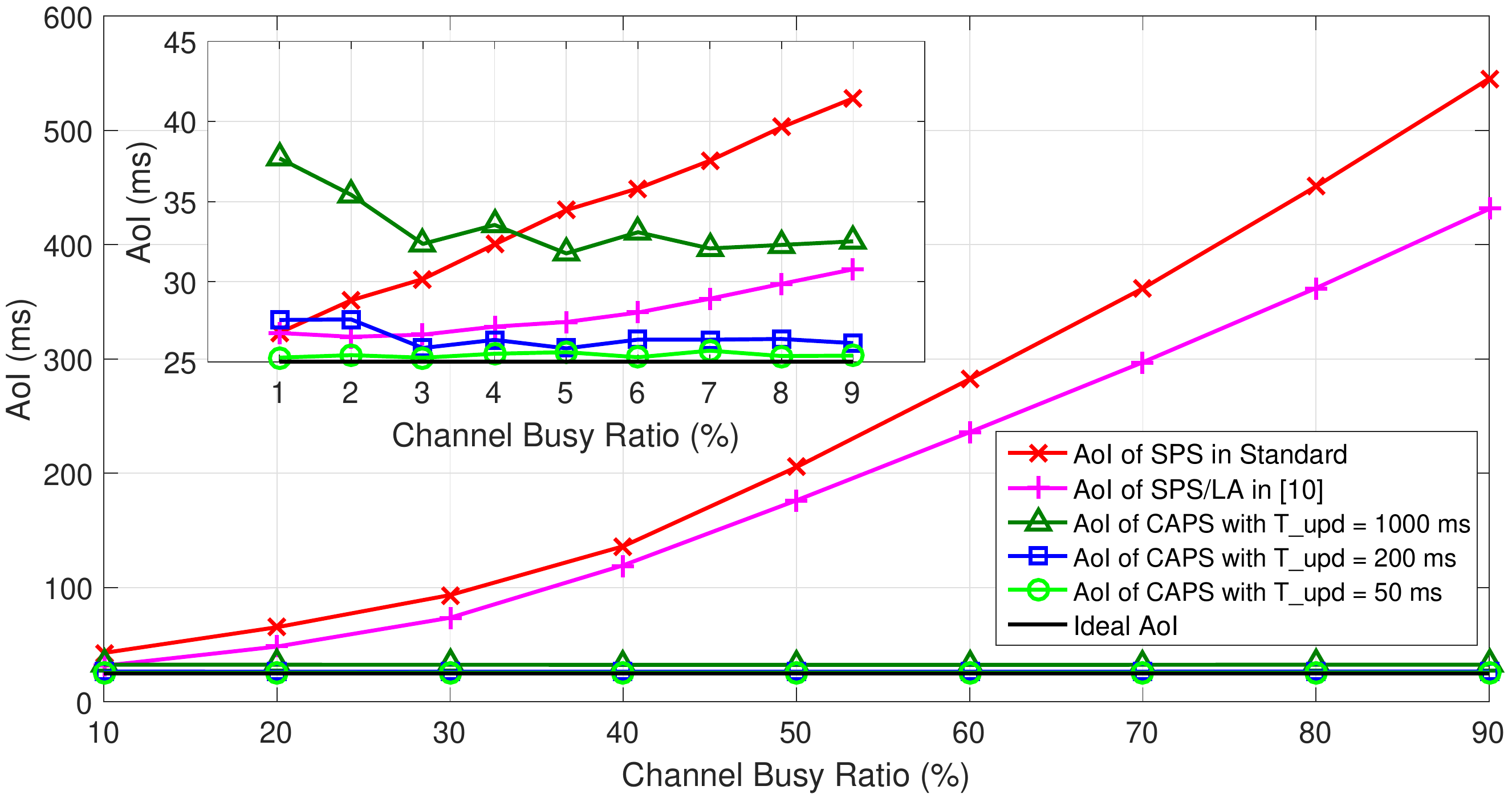}}
        \caption{Comparison of average AoI versus CBR among three different schemes.}
        \label{fig:aoi_comparison_three_schemes}
    \end{figure}

    Fig.\ref{fig:aoi_comparison_three_schemes} shows the comparison of average AoI versus CBR during the simulation that lasts for $100$ seconds. From Fig.\ref{fig:aoi_comparison_three_schemes}, the performance of CAPS scheme is obviously better than the other two schemes, where the curve of the average AoI of CAPS scheme is almost flat with CBR increasing while the performance of the other two schemes becomes worse with CBR increasing. When CBR is larger than $70\%$, the average AoI of the other two schemes is almost $10$ times as many as CAPS scheme. In addition, the average AoI of CAPS scheme is near the ideal AoI, where the ideal AoI has been mentioned in \ref{subsec:optimal_config}. From the perspective of $T_{\text{upd}}$, the better AoI is achieved with the smaller $T_{\text{upd}}$, which is the same as the result of the performance analysis in \ref{subsec:optimal_config}.
    
\begin{figure}[t]
        \centerline{\includegraphics[width = 0.82\columnwidth]{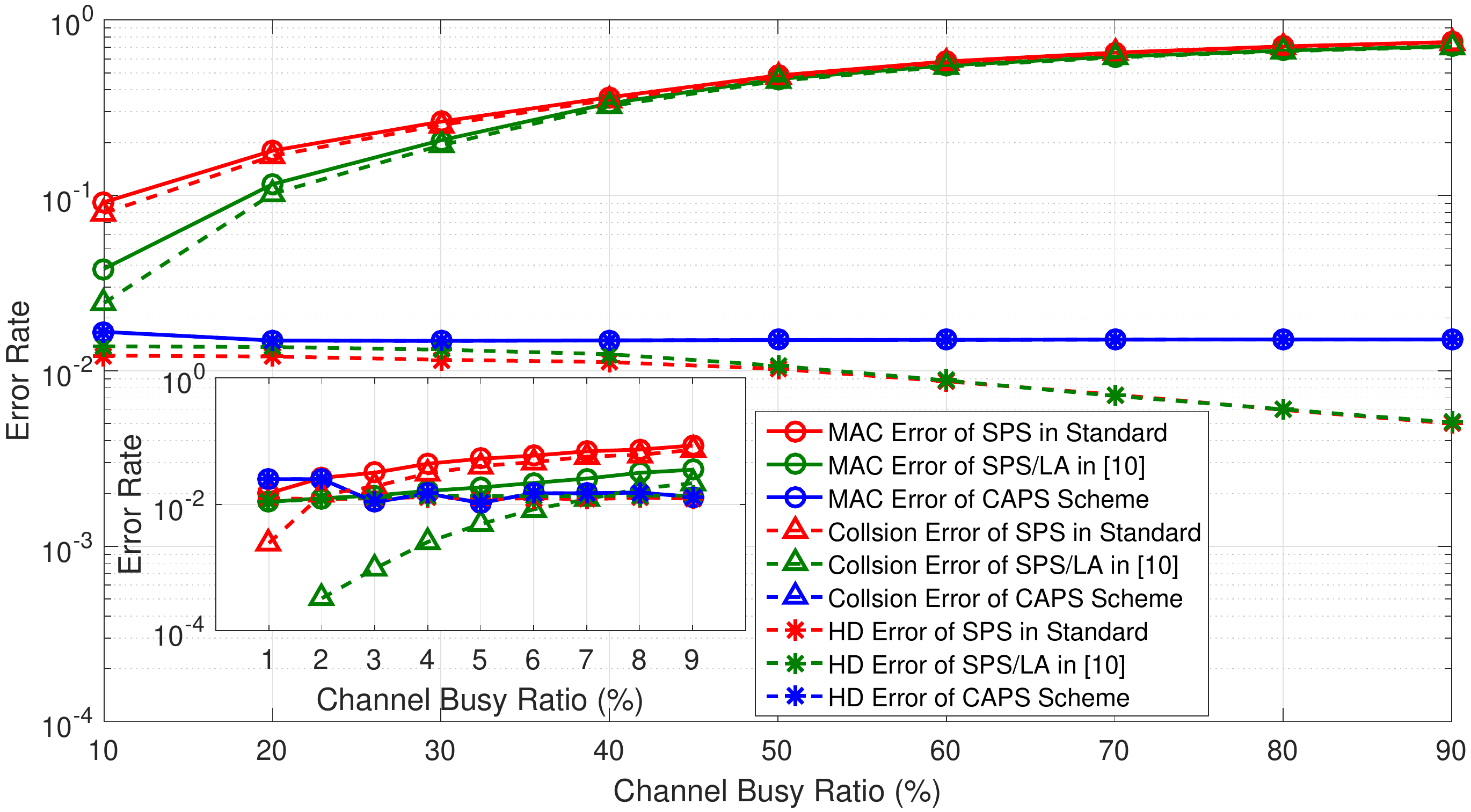}}
        \caption{Comparison of error rate versus CBR among three different schemes.}
        \label{fig:BLER_comparison_three_schemes}
    \end{figure}

    From the perspective of reliability, Fig. \ref{fig:BLER_comparison_three_schemes} shows comparisons of the error rate versus CBR when the simulation lasts for $100$ seconds. The MAC error here denotes the error caused by MAC layer operations, which contains both collision error and HD error. From Fig. \ref{fig:BLER_comparison_three_schemes}, the performance of CAPS scheme is obviously better than the other two schemes, and there is no error of collision when CBR $\leq 90\%$ in CAPS scheme. From the side, the results of reliability validate that AoI can also reflect reliability.
    
    \subsubsection{Comparison between Analysis and Simulation}
\begin{figure}
    \centering
    \subfigure[]{
    \begin{minipage}[c]{0.48\linewidth}
    \centering
    \includegraphics[width=7.9cm]{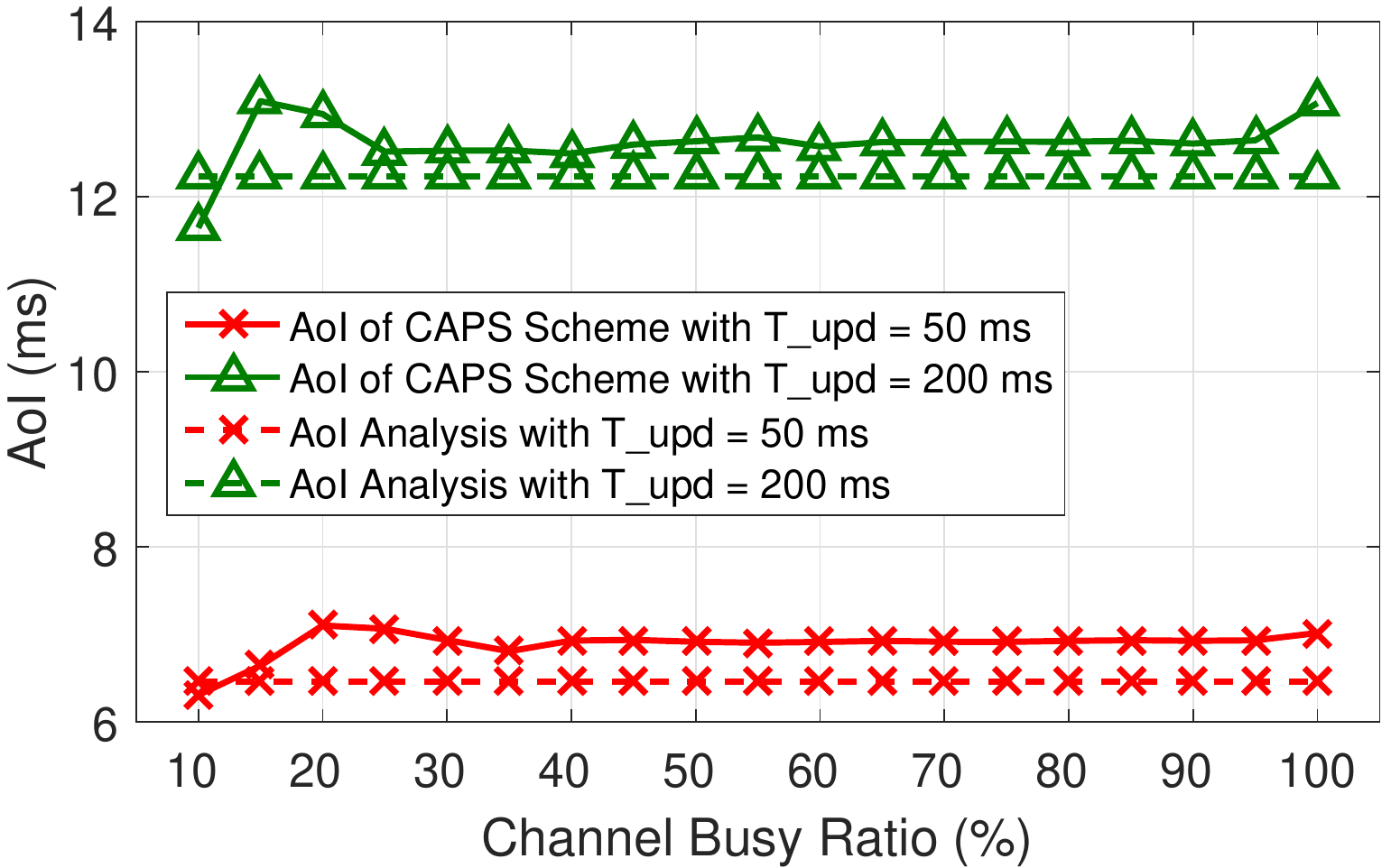}
    \label{fig:aoi_comparison_analysis_simulation_static}
    \vspace*{-10pt}
    \end{minipage}}
    \subfigure[]{
    \begin{minipage}[c]{0.48\linewidth}
    \centering
    \includegraphics[width=7.9cm]{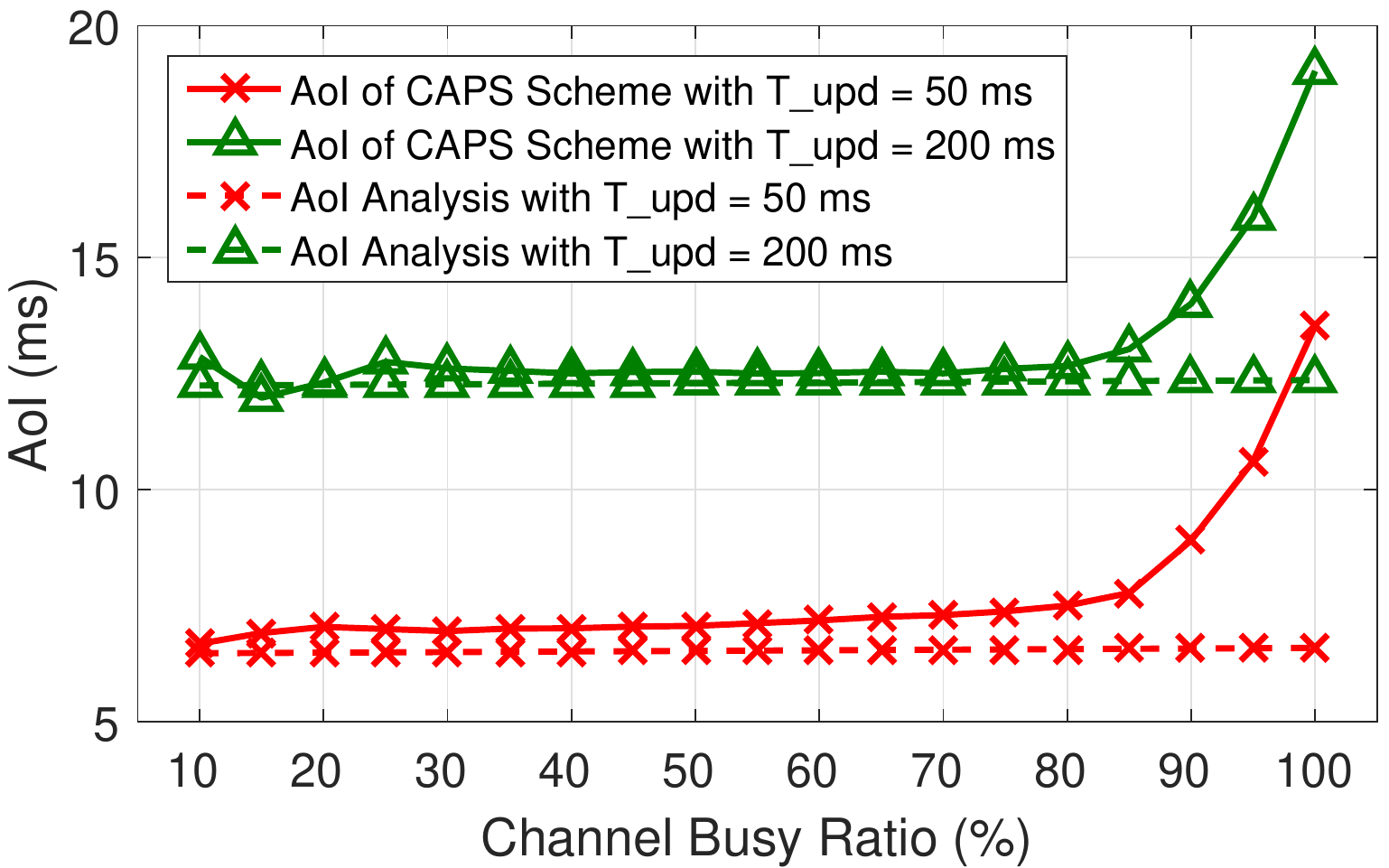}
    \label{fig:aoi_comparison_analysis_simulation_dynamic}
    \vspace*{-10pt}
    \end{minipage}}
    \vspace*{-5pt}
    \caption{(a) Comparison between theoretical analysis and simulation results in the scenario of static vehicular traffic flow. (b) Comparison between theoretical analysis and simulation results in the scenario of dynamic vehicular traffic flow, where $x=y=0.7$.}
    \end{figure}

    In order to validate the correctness of the result of the performance analysis, the results of analysis and simulation are compared in Fig.\ref{fig:aoi_comparison_analysis_simulation_static} and Fig.\ref{fig:aoi_comparison_analysis_simulation_dynamic}, where the RRI of all vehicles and $T_{\text{ost}}$ are equal to $10$ ms. Fig.\ref{fig:aoi_comparison_analysis_simulation_static} and Fig.\ref{fig:aoi_comparison_analysis_simulation_dynamic} correspond to the scenario of static and dynamic vehicular traffic flow respectively. From the comparison, it is obvious that the results of theoretical analysis are basically consistent with the simulation results while the simulation results get worse when CBR respectively overpasses $95\%$ and $80\%$ in the scenario of static and dynamic vehicular traffic flow. The deviation of analysis and simulation in very high CBR may be related to the convergence speed. If the convergence time is shortened, for example allowing vehicles to broadcast more sub-channels suspected of collision at the cost of increasing overhead, the deviation may be alleviated.
    
    \subsubsection{Verification of Optimal RRI and Optimization of the Range of CBR}
    \label{subsubsc_opt_RRI}
    As mentioned in \ref{subsec:optimal_config}, when $T_{\text{upd}}$ is fixed and $v \leqslant c$ is met, there is an optimal RRI. With the optimal RRI, the performance of AoI can achieve the minimum value. In order to validate the conclusion, a series of simulations are designed, where $T_{\text{ost}}$ is set to $10$ ms, $T_{\text{upd}}$ is set to $400$ ms and the scenarios of vehicular traffic flow consist of static and dynamic scenarios.
    
\begin{figure}
    \centering
    \subfigure[]{
    \begin{minipage}[c]{0.48\linewidth}
    \centering
    \includegraphics[width=7.9cm]{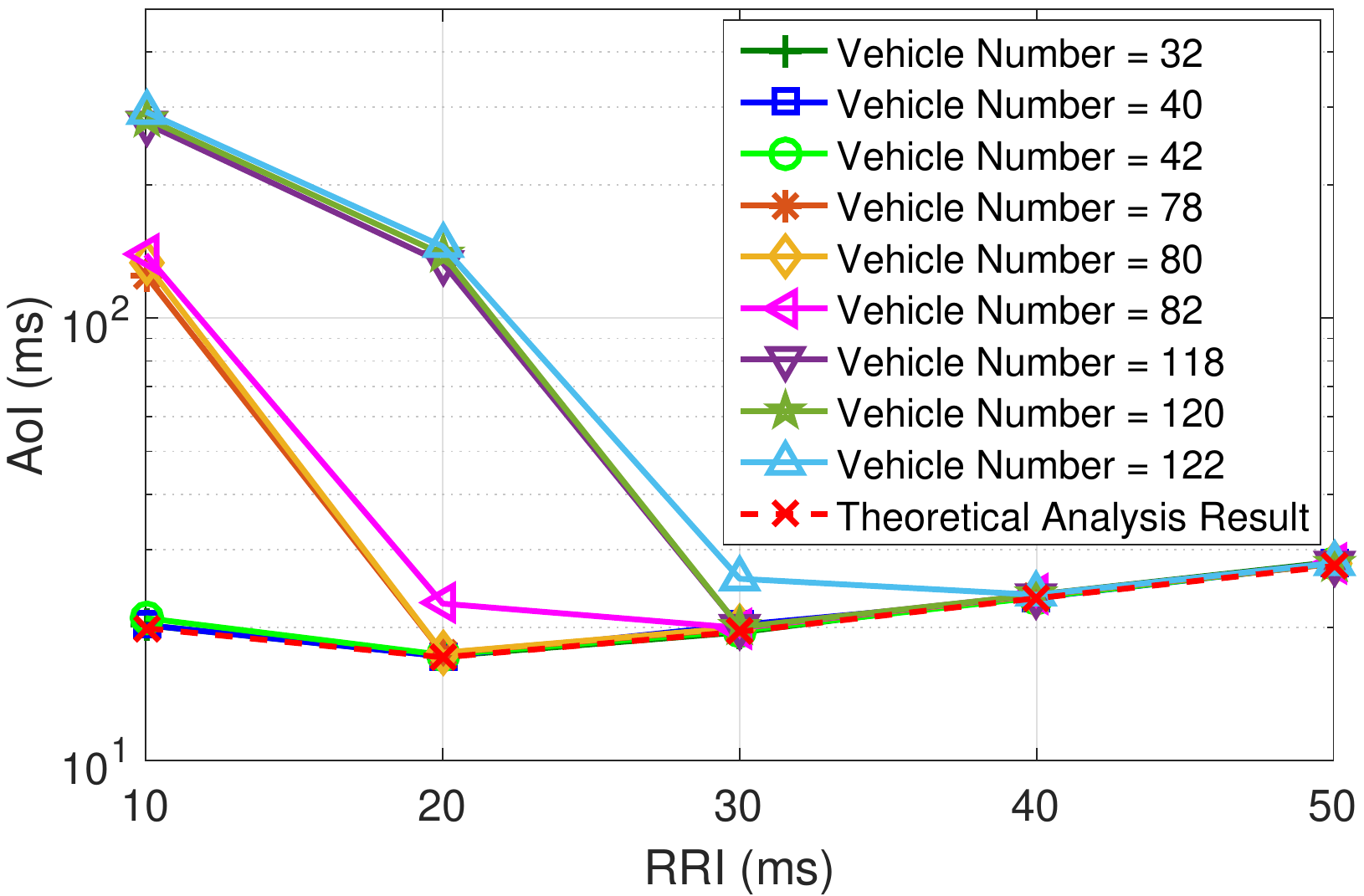}
    \label{fig:aoi_of_different_RRI_static}
    \vspace*{-10pt}
    \end{minipage}}
    \subfigure[]{
    \begin{minipage}[c]{0.48\linewidth}
    \centering
    \includegraphics[width=7.9cm]{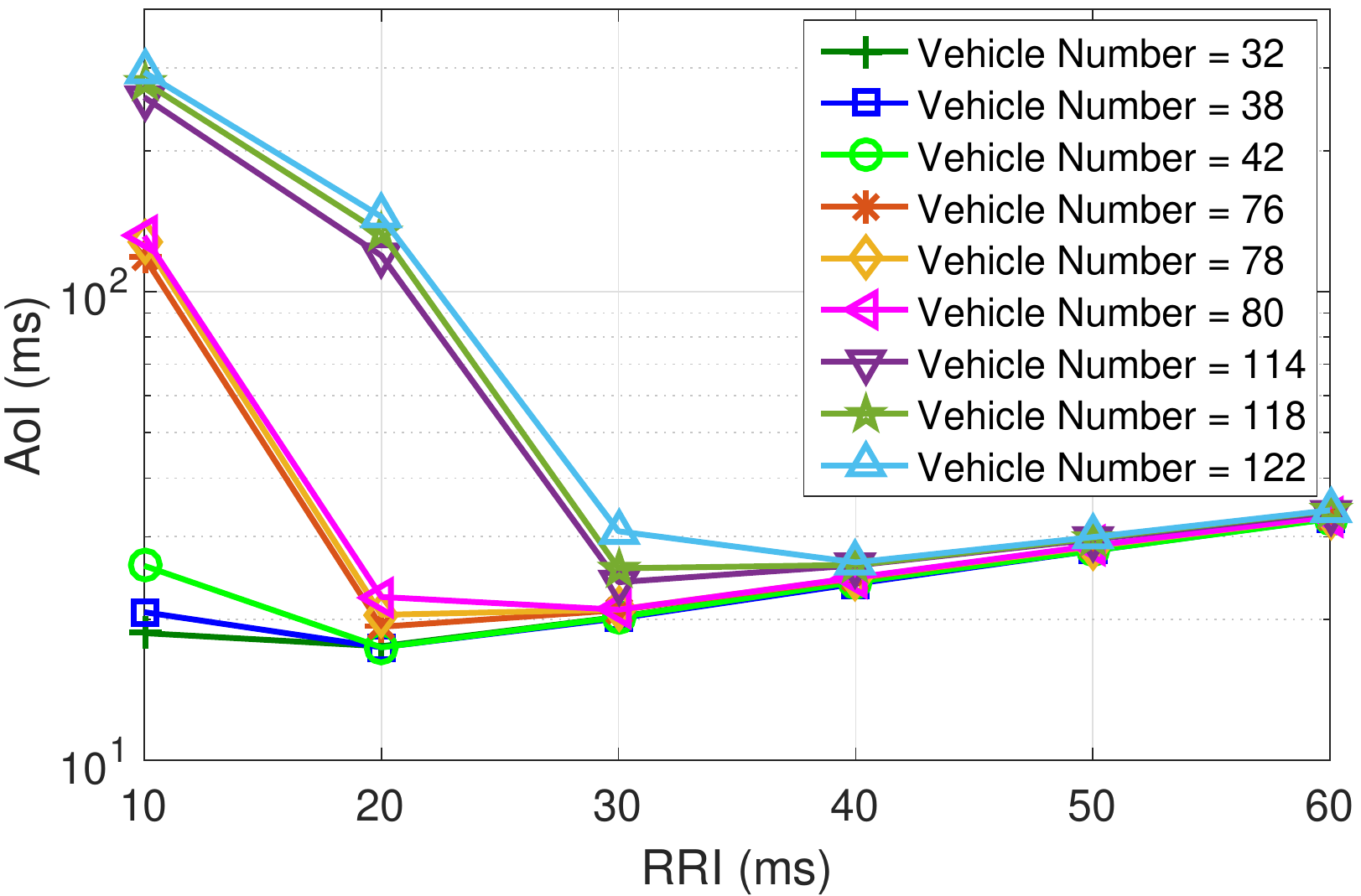}
    \label{fig:aoi_of_different_RRI_dynamic}
    \vspace*{-10pt}
    \end{minipage}}
    \vspace*{-5pt}
    \caption{(a) Average AoI versus RRI under different number of vehicles in the scenario of static vehicular traffic flow is shown and compared with the curve of the results of performance analysis with the same configuration. (b) Average AoI versus RRI under different number of vehicles in the scenario of dynamic vehicular traffic flow is shown, where $x=y=0.7$.}
    \end{figure}

    Fig.\ref{fig:aoi_of_different_RRI_static} shows the curve of the average AoI versus RRI under different number of vehicles in the scenario of static vehicular traffic flow. In the figure, dotted line indicates the result of theoretical analysis with the same configuration, which is corresponding to the result in Fig.\ref{fig:aoi_analysis_static}. 
    
    With different RRI, different number of vehicles corresponds respective CBR. Since the number of sub-channels in a sub-frame is equal to $4$, there are $40$ available sub-channels when the length of RRI is equal to $10$. The number of the available sub-channels with other RRI can be calculated in the same manner as well. From Fig.\ref{fig:aoi_of_different_RRI_static}, it is obvious that when the number of vehicles equals to $32$ and $40$, the curves of them nearly accord with the result of theoretical analysis. Once the number of vehicles overpasses the number of available sub-channels under a certain RRI, i.e. $v>c$, the performance will be significantly inferior to the result of theoretical analysis. Therefore, based on the observation on Fig.\ref{fig:aoi_of_different_RRI_static}, it is concluded that in the scenario of static vehicular traffic flow, only when the CBR is under $100\%$ with vehicles using the optimal RRI of theoretical analysis, the optimal RRI can be adopted. Otherwise, the RRI used by the vehicles should be the shortest available RRI that can let CBR under $100\%$. In order to implement the conclusion in the practical scenarios, the simplest idea is to dynamically adjust respective RRI through estimating the CBR in real time, which will be mentioned in \ref{subsubsec_adaptive_RRI}.
    
    From the perspective of dynamic scenario, the simulation results are shown in Fig.\ref{fig:aoi_of_different_RRI_dynamic}, where $x=y=0.7$. Since the results of theoretical analysis are different for different number of vehicles in the scenario of dynamic vehicular traffic flow, the curves of theoretical analysis are omitted, which will not influence the analysis on the results.

    In Fig.\ref{fig:aoi_of_different_RRI_dynamic}, in addition to the condition that $v>c$, when CBR is under but near $100 \%$, the performance will also become worse. According to the result of theoretical analysis, the optimal RRI should be $20$ ms. However, only when the number of vehicles is under $78$, where the CBR approximately equals to $98\%$ with RRI of $20$ ms, the RRI is optimal. For the curve of the vehicular number of $80$, $114$ and $118$, the optimal RRI is $30$ ms, where all of their CBRs are under $98.3\%$ as well. The optimal RRI of the vehicular number of $122$ equals to $40$ ms, where the CBR does not overpass $98\%$.
    
    Different from the static scenario, through comparing the relation between CBR and the optimal RRI among all curves, it is concluded that in the scenario of dynamic vehicular traffic flow, where $x=y=0.7$, only when the CBR is under $98\%$ with vehicles using the optimal RRI of theoretical analysis, the optimal RRI can be adopted. Otherwise, the RRI used by the vehicles should be the shortest available RRI that can let CBR under $98\%$. With the higher dynamic rate, the threshold will be lower. In addition, for implementing the conclusion in the practical scenarios, besides the process of CBR sensing, the sensing of dynamic rate is also necessary, for example through identifying the vehicle ID, which will be studied in the future work.
    
    \subsection{Simulations in the Freeway Scenario}
    \label{subsec_sim_freeway}
    Through the simulations in \ref{subsec_sim_mac}, it is shown that CAPS has many advantages over the state-of-the-art on the MAC layer. However, in the practical scenarios, the performance of CAPS may be influenced by the wireless channel and the mobility of vehicles. Therefore, it is necessary to simulate CAPS scheme in a more practical scenario. In this part, the mobility of vehicles, physical channel and the loss of physical layer process are considered in the simulation. In addition, the problems that may occur in the real conditions, e.g., false detection and missing detection of packet collisions, are analyzed as well. 
    
    \subsubsection{Models and Configurations of the Simulations}
    According to 3GPP 36.885 \cite{3gpp:36.885}, the freeway case in the mobility model is adopted, where the corresponding channel model is the line-of-sight (LOS) case of WINNER+ B1 \cite{meinila2009winner}. The detailed parameters are shown in Table. \ref{freeway_configurations}. In addition, the loss of physical layer process is obtained from the results of the link level performance in \cite{hu2017link}. In the simulations of this part, the configuration of the message size is that one 300-byte message followed by four 190-byte messages, which is from 3GPP 36.885 \cite{3gpp:36.885} as well. For the resource occupancy, the 190-byte and the 300-byte messages respectively require one and two sub-channels. The other necessary configurations not mentioned in Table. \ref{freeway_configurations} keep the same designs as Table. \ref{Simulation Settings} except the traffic flow of vehicles, which is replaced by the freeway scenario.
    
    \begin{table} [t] 
    \centering 
    \caption{Configurations in the Freeway Scenarios} 
    \label{freeway_configurations}
    \footnotesize
    \begin{tabular}{c c|c c}  
    \hline
    \textbf{Parameter}&\textbf{Value}&\textbf{Parameter}&\textbf{Value}\\
    \hline
    Number of Vehicles ($v$) & $50/150/250$ & Resource Offset Update Period ($T_{\text{upd}}$) & $200$ ms \\
    \hline
    Freeway Length & $2$ km & Absolute Vehicle Speed & $70/140$ km/h \\
    \hline
    Message Size & $190/300$ bytes & Number of Lanes & $6$ in dual directions \\
    \hline
    Lane Width & $4$ m & Shadowing Distribution & log-normal \\
    \hline
    Shadowing Standard Deviation & $3$ dB & Transmission Power & $23$ dBm \\
    \hline
    Working Frequency & $5.9$ GHz & Threshold for Collision Detection\footnotemark[1] & $-95$ dBm \\
    \hline
    \end{tabular}  
    \end{table} 
    \footnotetext[1]{The threshold is one of the conditions that is used to judge whether a sub-channel is suspected of collision as mentioned in Definition \ref{defi_col_sus}.}
    
    \subsubsection{Analysis of Results}
\begin{figure}
    \centering
    \subfigure[]{
    \begin{minipage}[c]{0.48\linewidth}
    \centering
    \includegraphics[width=7.9cm]{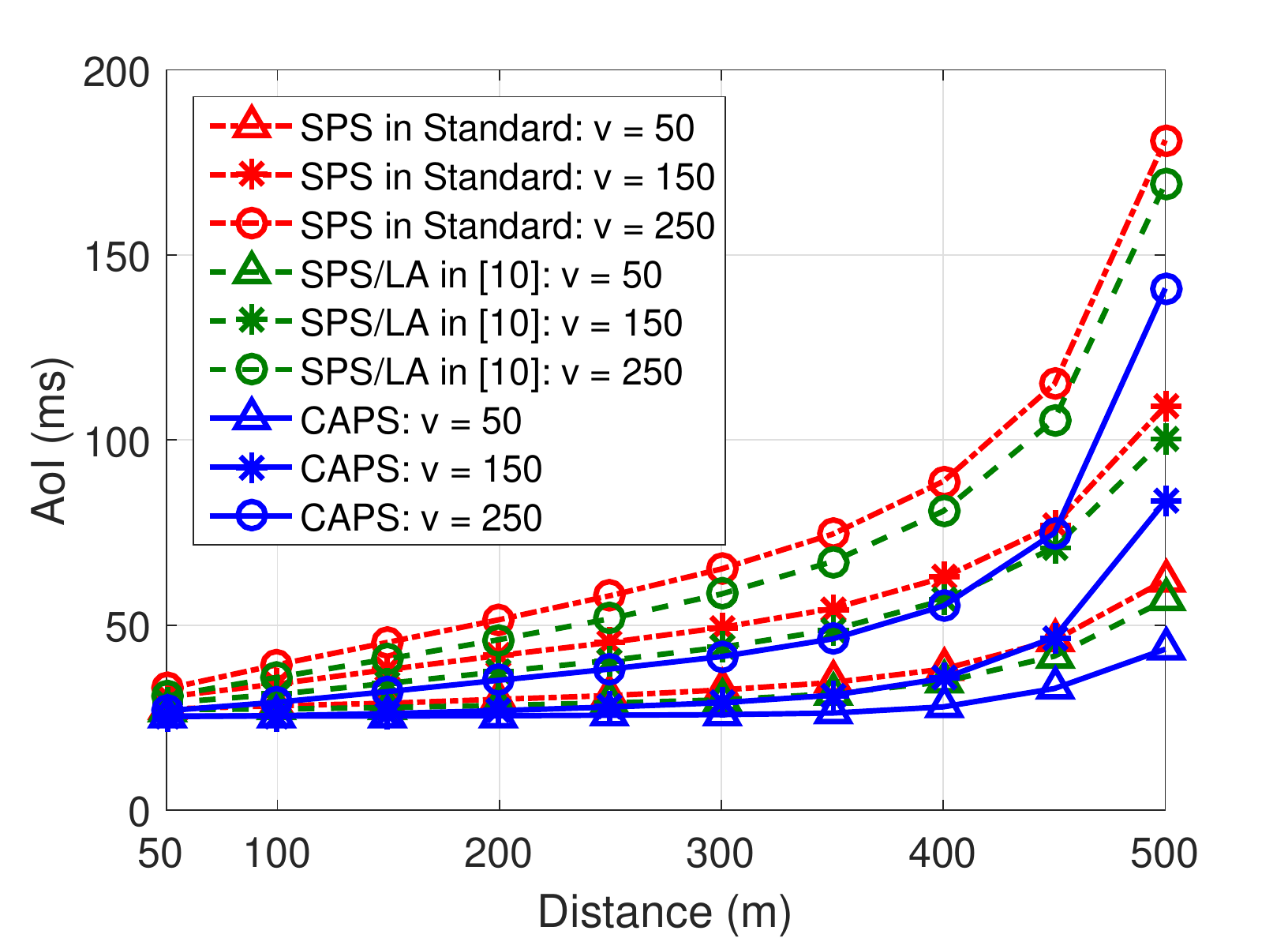}
    \vspace*{-10pt}
    \end{minipage}}
    \subfigure[]{
    \begin{minipage}[c]{0.48\linewidth}
    \centering
    \includegraphics[width=7.9cm]{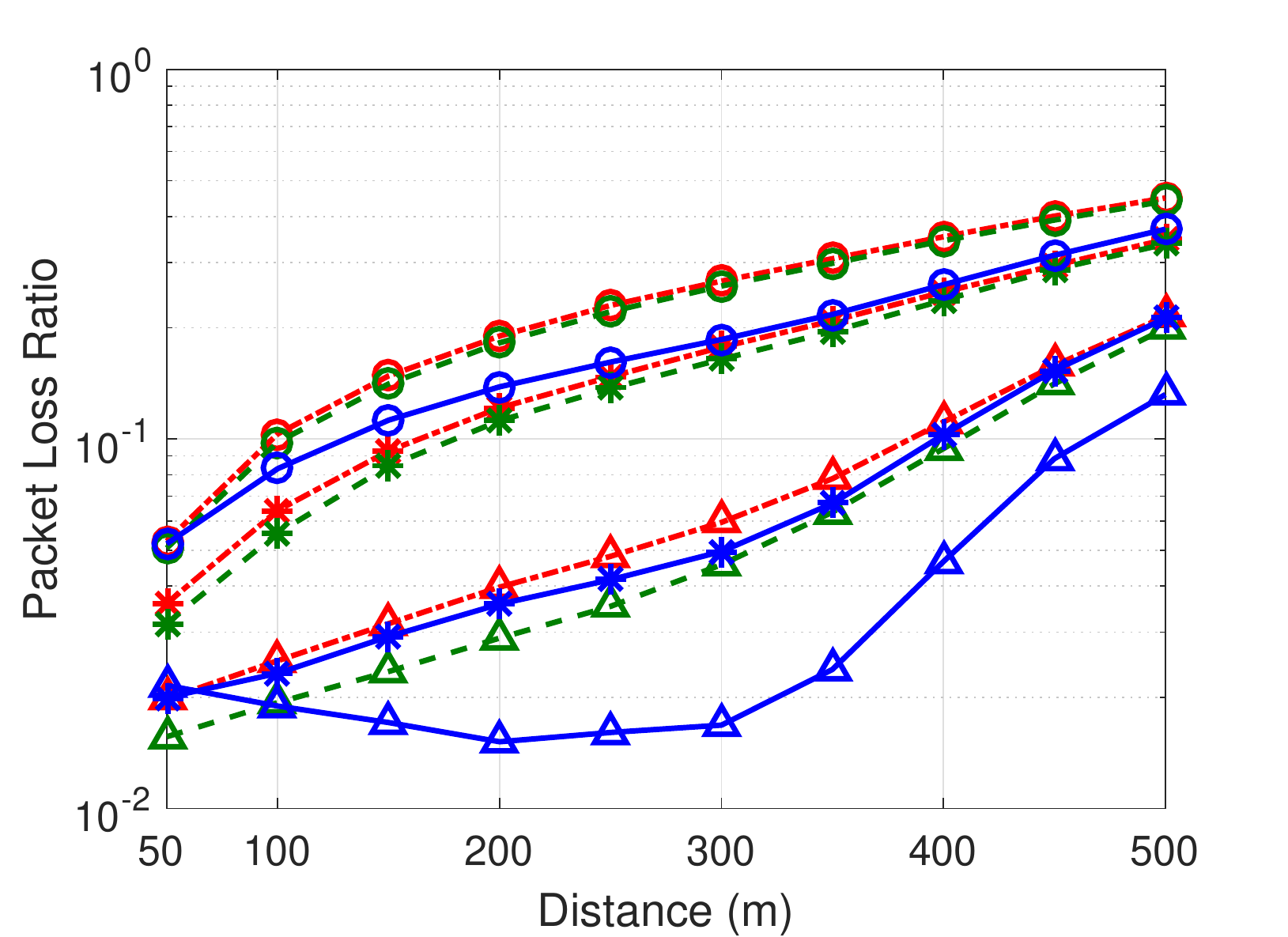}
    \vspace*{-10pt}
    \end{minipage}}
    \subfigure[]{
    \begin{minipage}[c]{0.48\linewidth}
    \centering
    \includegraphics[width=7.9cm]{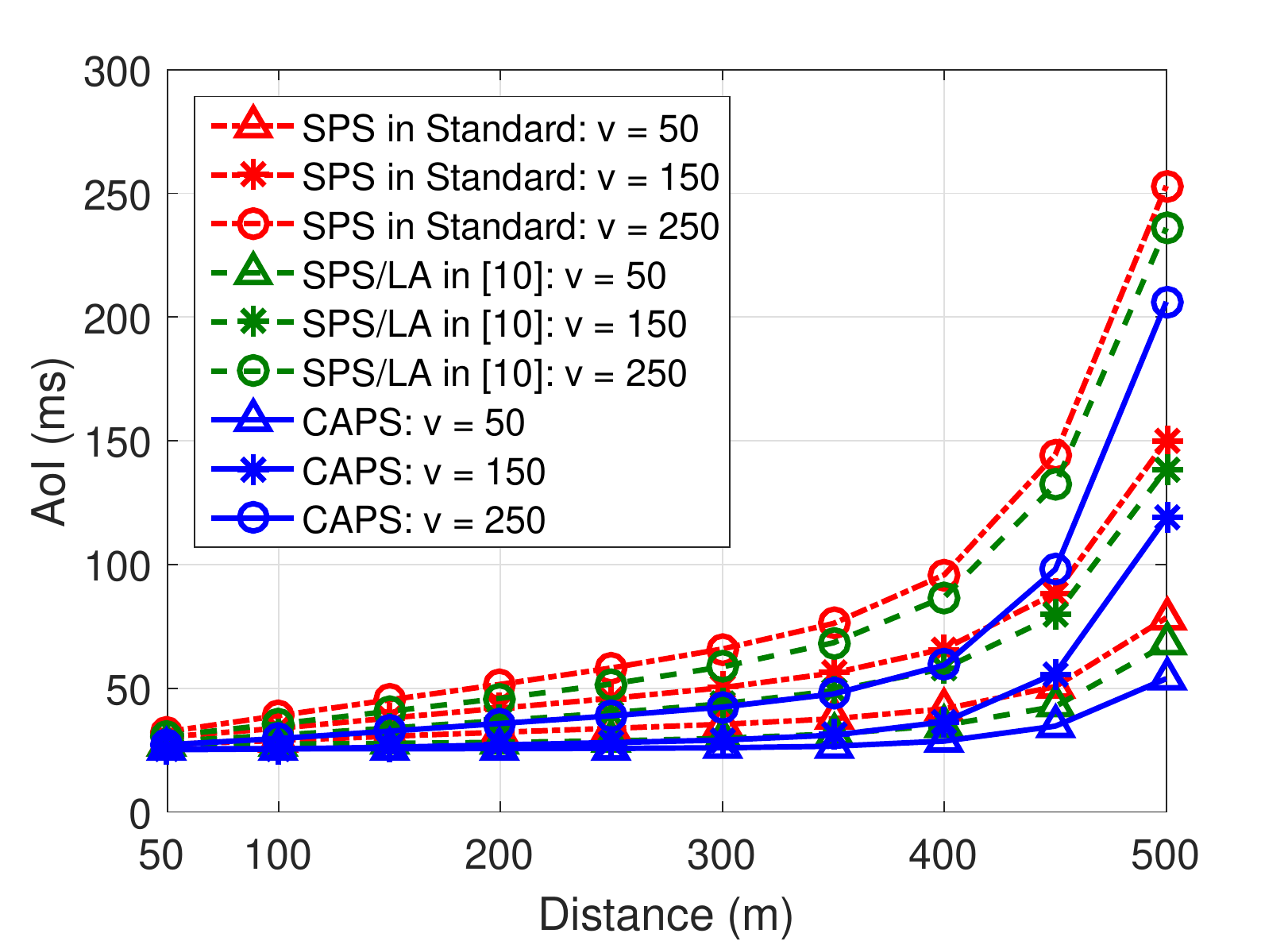}
    \vspace*{-10pt}
    \end{minipage}}
    \subfigure[]{
    \begin{minipage}[c]{0.48\linewidth}
    \centering
    \includegraphics[width=7.9cm]{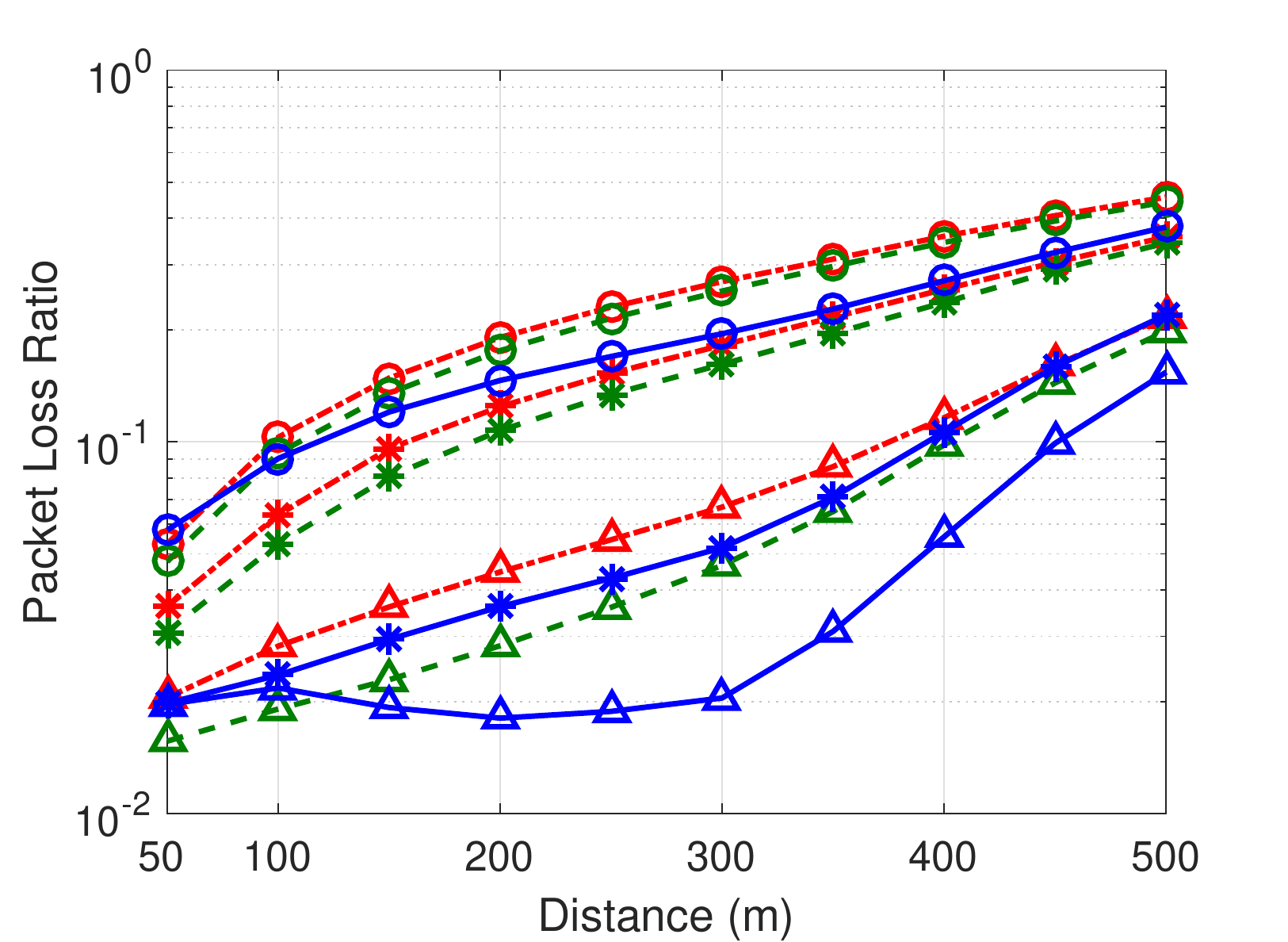}
    \vspace*{-10pt}
    \end{minipage}}
    \vspace*{-5pt}
    \caption{(a) Comparison of average AoI versus distance among three different schemes with the absolute speed of $70$ km/h. (b) Comparison of packet loss ratio with the absolute speed of $70$ km/h. (c) Comparison of average AoI with the absolute speed of $140$ km/h. (d) Comparison of packet loss ratio with the absolute speed of $140$ km/h.}
    \label{fig_comparison_freeway_three_schemes}
    \end{figure}

    After the physical channel effect is introduced, the simulation results are shown in Fig. \ref{fig_comparison_freeway_three_schemes}. As shown in Fig. \ref{fig_comparison_freeway_three_schemes}, the AoI and reliability performance of CAPS are still better than other two schemes with different vehicle number $v$ and absolute speed.
    
    In CAPS, since the packet collisions are detected by the receivers, when collaboration message is broadcasted, the related vehicles with packet collisions may not hear each other due to the communication range, which is always considered as the hidden terminal problem. Therefore, CAPS can mitigate the hidden terminal problems to some extent. To evaluate the ability, the ratio $r_\text{ht}$ of the number of the solved hidden terminal problems $n_\text{ht}$ among the number of the solved collisions $n_\text{col}$ is counted in Table. \ref{tab_hidden_terminal}. As shown in Table. \ref{tab_hidden_terminal}, the average ratio $r_\text{ht}$ is about $44.6\%$, which proves the ability to mitigate the hidden terminal problems.
    
    \begin{table} [t] 
    \centering 
    \caption{Ability to Mitigate the Hidden Terminal Problems} 
    \label{tab_hidden_terminal}
    \footnotesize
    \begin{tabular}{c|c c c c c c}  
    \hline
    \tabincell{c}{\textbf{Configuration}\\ \textbf{(vehicle number, absolute speed)}}& (50, 70) & (50, 140) & (150, 70) & (150, 140) & (250, 70) & (250, 140) \\
    \hline
    \textbf{$n_\text{ht}$} & $11$ & $16$ & $3266$ & $3935$ & $18296$ & $19159$ \\
    \hline
    \textbf{$n_\text{col}$} & $27$ & $30$ & $5890$ & $6965$ & $59190$ & $62148$  \\
    \hline
    \textbf{$r_\text{ht}$} & $40.7\%$ & $53.3\%$ & $55.5\%$ & $56.5\%$ & $30.9\%$ & $30.8\%$  \\
    \hline
    \end{tabular}  
    \end{table}
    
    In the practical scenarios, affected by the wireless channel, the vehicles using CAPS scheme maybe not detect the packet collisions correctly all the time, where false detection and missing detection of packet collisions may occur. 
    
    False detection always results from the signal attenuation and propagation effects. When the message with no collision is not correctly decoded due to the wireless channel and the energy of the message is higher than the threshold for collision detection, the corresponding sub-channel will be considered to be suspected of collision and then false detection will occur. To mitigate false detection, it is necessary to select a suitable threshold for collision detection. Through extensive simulations, Table. \ref{tab_false_detection} is obtained, which provides the ratio $r_\text{fd}$ of the number of false detection $n_\text{fd}$ among the total number of collision detection $n_\text{cd}$. As shown in Table. \ref{tab_false_detection}, the threshold for collision detection is required to be high enough to avoid false detection. However, the higher the threshold is, the less collisions the vehicles can detect, which leads to missing detection and is reverse to the aim of CAPS. Therefore, $-95$ dBm can be chosen to be the threshold value according to Table. \ref{tab_false_detection}.
    
    \begin{table} [t] 
    \centering 
    \caption{Ratio of False Detection}
    \label{tab_false_detection}
    \footnotesize
    \begin{tabular}{c|c c c c c c c c}
    \hline
    \tabincell{c}{\textbf{Threshold for Collision} \\\textbf{Detection (dBm)}} & $-102$ & $-101$ & $-100$ & $-99$ & $-98$ & $-97$ & $-96$ & $-95$  \\
    \hline
    \textbf{$n_\text{fd}$} & $17061$ & $7786$ & $2473$ & $540$ & $79$ & $11$ & $2$ & $0$  \\
    \hline
    \textbf{$n_\text{cd}$} & $17513$ & $8123$ & $2675$ & $677$ & $121$ & $46$ & $31$ & $49$  \\
    \hline
    \textbf{$r_\text{fd}$} & $97.4\%$ & $95.9\%$ & $92.5\%$ & $79.8\%$ & $65.3\%$ & $23.9\%$ & $6.5\%$ & $0.0\%$  \\
    \hline
    \end{tabular}
    \end{table}
    
    For missing detection, in addition to the threshold for collision detection, near-far effect is the other reason that the vehicles neglect the collisions. For example, there are two vehicles transmitting in the same sub-channel and both of them are in the communication range of the receiving vehicle. One of the transmitting vehicles is close to the receiver and the other is a little farther. If the signal to interference plus noise ratio (SINR) of the closer vehicle is high enough, the receiver may neglect the packet collision. To evaluate the risk from near-far effect, the average distance $d_\text{farther}$ between the receiver and the farther vehicle is calculated through the simulations, which is shown in Table. \ref{tab_missing_detection}. It is noted that the farther vehicles mentioned in Table. \ref{tab_missing_detection} are still in the communication range of the receiver. According to the results in the table, the average distance of the farther vehicle is about $397$ m, which is far enough for the general use cases. 
    
    \begin{table} [t] 
    \centering 
    \caption{Average Distance of the Farther Vehicle When Missing Detection Occurs}
    \label{tab_missing_detection}
    \footnotesize
    \begin{tabular}{c|c c c c c c}
    \hline
    \tabincell{c}{\textbf{Configuration}\\ \textbf{(vehicle number, absolute speed)}}& (50, 70) & (50, 140) & (150, 70) & (150, 140) & (250, 70) & (250, 140) \\
    \hline
    \textbf{$d_\text{farther}$ (m)} & $405$ & $442$ & $416$ & $414$ & $355$ & $351$  \\
    \hline
    \end{tabular}  
    \end{table}
    
    In summary, including the hidden terminal problems, false detection and missing detection, the CAPS scheme still presents a better performance than the other two schemes in Fig. \ref{fig_comparison_freeway_three_schemes}.
    
    \subsubsection{Adaptive RRI}
    \label{subsubsec_adaptive_RRI}
\begin{figure}
    \centering
    \subfigure[]{
    \begin{minipage}[c]{0.48\linewidth}
    \centering
    \includegraphics[width=7.9cm]{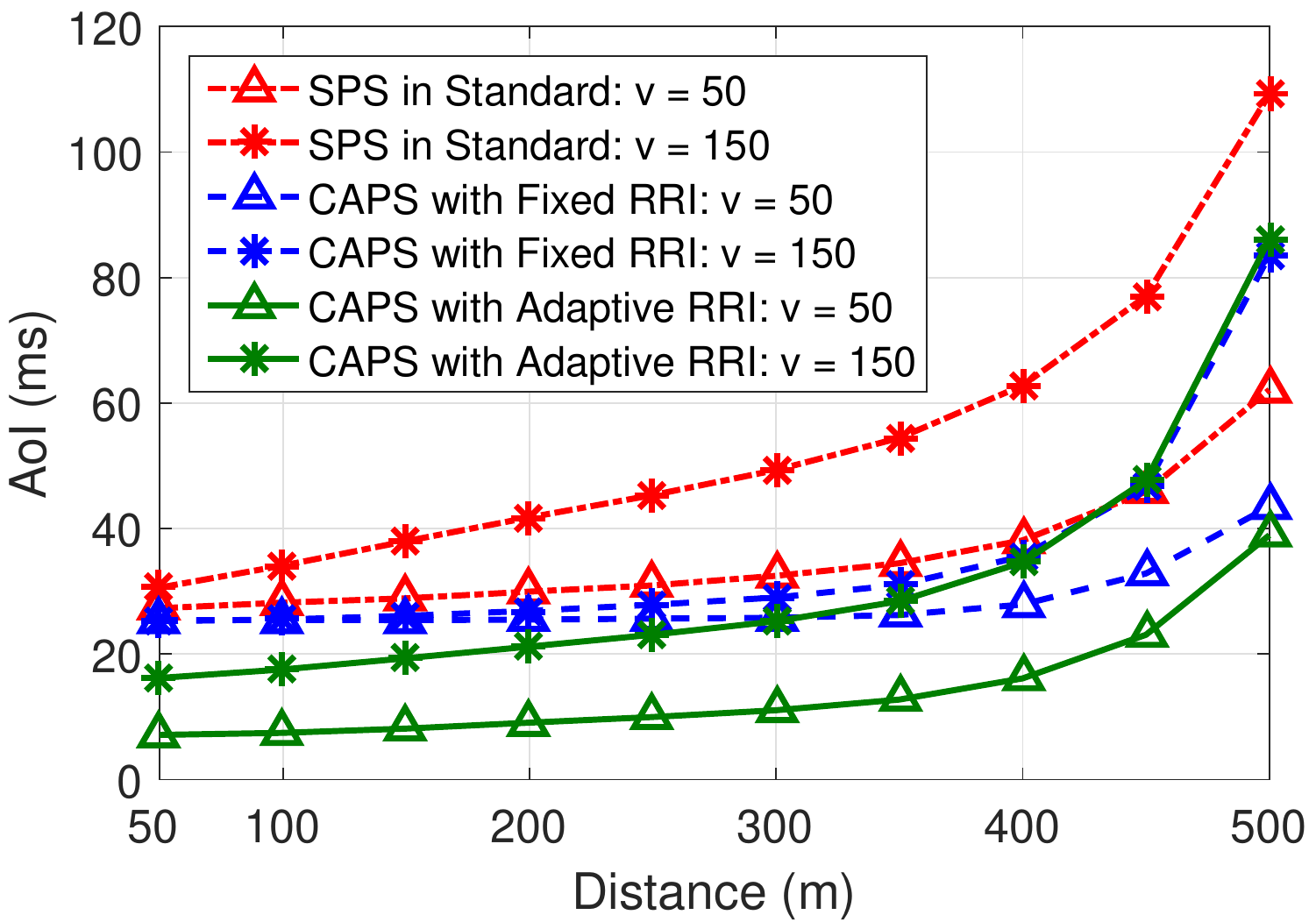}
    \vspace*{-10pt}
    \end{minipage}}
    \subfigure[]{
    \begin{minipage}[c]{0.48\linewidth}
    \centering
    \includegraphics[width=7.9cm]{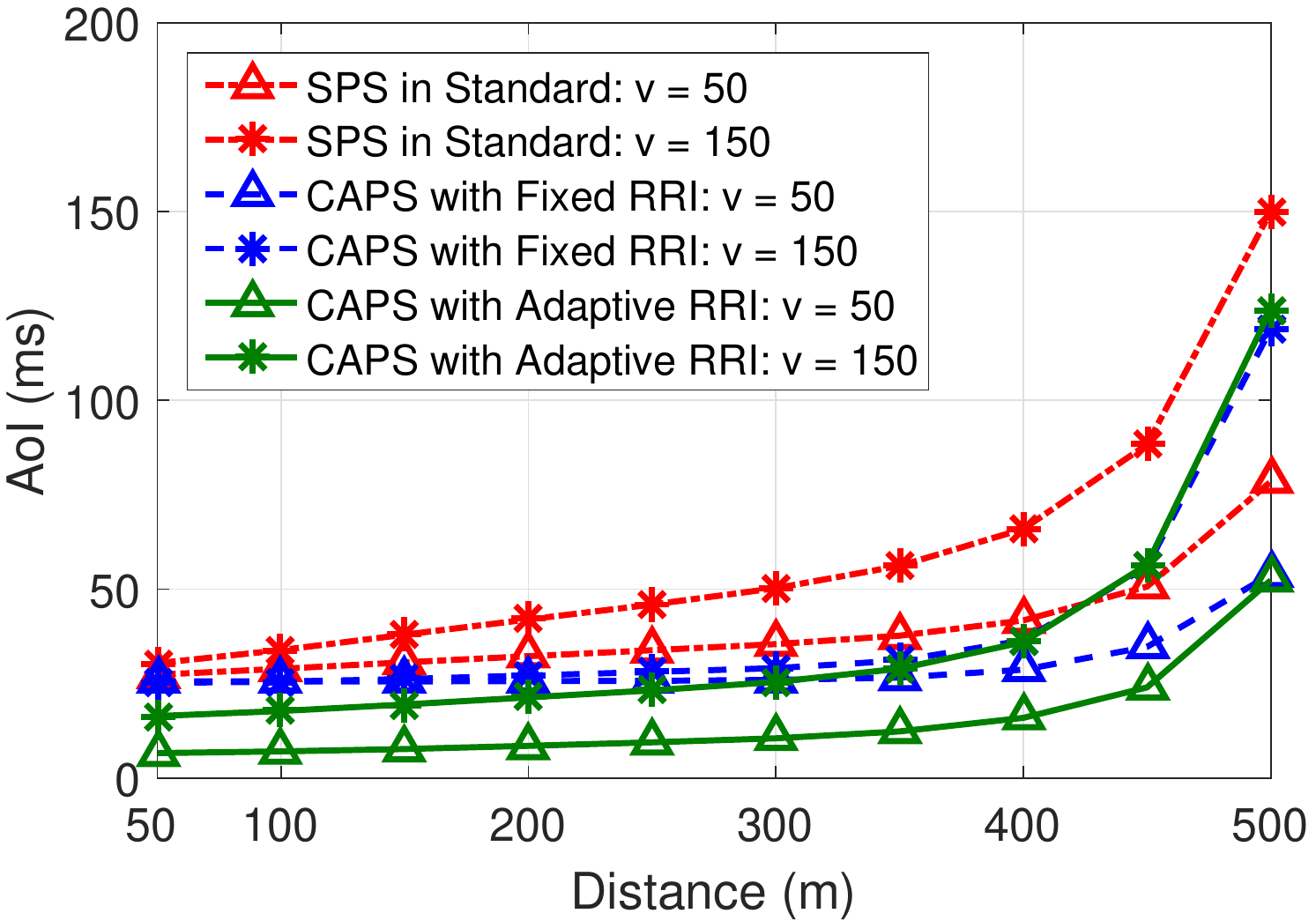}
    \vspace*{-10pt}
    \end{minipage}}
    \vspace*{-5pt}
    \caption{(a) Comparison of average AoI versus distance with the absolute speed of $70$ km/h. (b) Comparison of average AoI with the absolute speed of $140$ km/h.}
    \label{fig_comparison_adaptive_RRI}
    \end{figure}

\begin{figure}[t]
        \centerline{\includegraphics[width = 0.92\columnwidth]{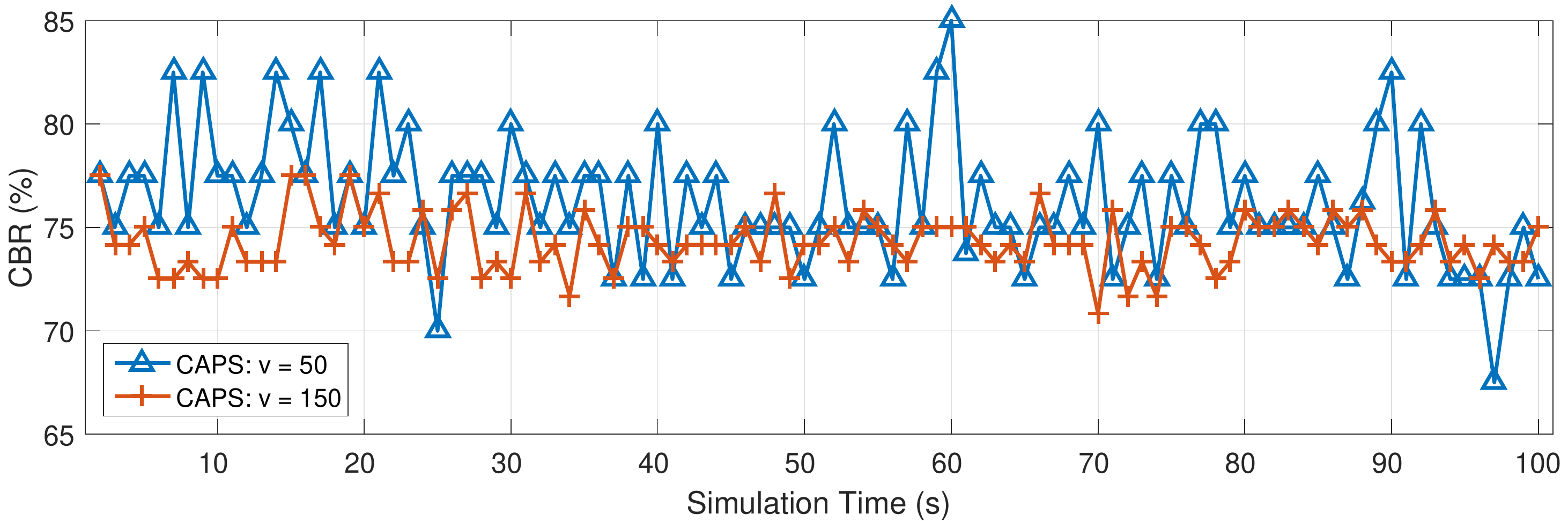}}
        \caption{CBR variation during the simulations with different number of vehicles.}
        \label{fig_CBR_variation}
    \end{figure}

    From the comparison and analysis above, CAPS scheme with the fixed RRI has shown a better performance than the standard SPS scheme and the state-of-the-art scheme. In this part, adaptive RRI is adopted in CAPS to further enhance the AoI performance according to the theoretical results in \ref{subsec:optimal_config} and the simulation analysis in \ref{subsubsc_opt_RRI}. To be specific, in the simulation of adaptive RRI, the vehicles would automatically adjust their own RRI depending on the current CBR. When CBR is lower than $40 \%$, the vehicles would use a shorter RRI. When CBR is higher than $80 \%$, the vehicles would use a longer RRI. The available RRI should be equal or larger than the optimal RRI calculated in Appendix \ref{sec_static} and meet $\alpha_{\text{RRI}} = mT_\text{ost}, m \in \mathbb{Z}^{+}$. As shown in Fig. \ref{fig_comparison_adaptive_RRI}, compared with the version of fixed RRI, the performance of CAPS with adaptive RRI is further enhanced. In addition, with the adaptive RRI mechanism, the real-time CBR can be controlled in the range from $65 \%$ to $85 \%$, which is shown in Fig. \ref{fig_CBR_variation}. Therefore, the condition $v \leqslant c$ can be met in CAPS in the practical scenarios.
    
    \section{Conclusions and Future Work}
    \label{sec_conc}
    In this paper, we have proposed a distributed resource allocation scheme, i.e., CAPS, by piggybacking the collaboration messages for collision avoidance with acceptable signaling overhead. Meanwhile, AoI is introduced as the metric to evaluate the performance of information timeliness in the scenario of vehicular direct communications. Based on theoretical analyses, the convergence of CAPS is proved. It is found that CAPS scales well with the vehicle population, as the average AoI is merely slightly affected by that. Additionally, the optimal RRI is derived in closed-form, under a certain number of vehicles. According to the simulation results in the MAC layer, the performance of CAPS is significantly better than the baseline schemes, including the standardized scheme and state-of-the-art enhancements. Specifically, when CBR is larger than $70\%$, the average AoI of CAPS is only $10\%$ of the other baselines. Furthermore, the average AoI of CAPS scheme is shown to be near-optimal, by comparing the simulation results to the theoretical optimum derived before. The optimal RRI to optimize AoI is also derived, under a wide range of CBR for both static and dynamic vehicular traffic flow scenarios. On the other hand, based on the simulation results in the freeway scenario, CAPS still shows a better performance of AoI and reliability. Meanwhile, the more practical problems, i.e., hidden terminal problems, false detection and missing detection, have been analyzed, which explains the feasibility of the proposed scheme in the practical scenarios. With the adaptive RRI mechanism, the paper provides a method to adjust the transmission interval in the real time, which further enhances the AoI performance of CAPS.
    
    From the perspective of application scenarios, we note that our proposed scheme is not only suitable for vehicular direct communications but also other scenarios of distributed periodical transmissions such as sensor sharing networks and emergency voice communication networks without cellular coverage. Essentially, CAPS can achieve good performance with persistent periodic packet transmissions, when vehicles have enough time to adapt their resource allocation. However, it may not be applicable to scenarios where a large number of aperiodic packets need to be transmitted in a short period of time, because collaboration becomes useless when packets are transmitted aperiodically. In practical V2X scenarios, it is certain that both periodic packets, e.g., status messages, and aperiodic packets, e.g., emergency messages, are required. Therefore, designing a scheme that can meet the transmission needs of both periodic and aperiodic messages requires future work.
    
    
    \bibliographystyle{ieeetr}
    \bibliography{spatial_aoi}
    
    \clearpage
    \begin{center}
	\textbf{\huge Supplementary Materials}    
	\end{center}
    
    \appendices
    \section{Convergence Analysis}
    \label{sec_proof}
    In the situation of static vehicular traffic flow, the aim of our scheme is to let the channel occupancy reach a stable state, in which there is no packet collision. However, the re-selection process may lead to additional packet collisions, which may make channel occupancy unable to converge to a stable state, even result in more packet collisions. Therefore, the convergence of the scheme will be analyzed in this section.
    
    When $v$ vehicles randomly select resources among $c$ sub-channels and each vehicle only select one sub-channel to transmit data, there is a certain probability of packet collision. In order to calculate the number of packets in each sub-channel after selection, we define $p_{i,j,c,v}$ as the probability that there are $i$ sub-channels with $j$ packets when $v$ vehicles randomly select resources among $c$ sub-channels, where $v \geq j \geq 0$. Therefore, $p_{i,j,c,v}$ should satisfy $\sum_{i=0}^{c} p_{i,j,c,v} = 1$.
    
    When $j=1$,
    \begin{iarray}
    \label{equ:p_j_1}
        p_{i,1,c,v} &=& {\binom{c}{i}} {\binom{v}{i}} \left(\frac{i}{c} \right)^{i} \left(\frac{c-i}{c} \right)^{v-i} \left(\frac{1}{i} \right)^{i} i! \left(1 - \sum_{k=1}^{c - i} p_{k,1,c - i,v - i} \right), \nonumber\\
        &=& {\binom{c}{i}} {\binom{v}{i}}  \left(1 - \frac{i}{c} \right)^{v - i}  \left(\frac{1}{c} \right)^{i}  i!  \left(1 - \sum_{k=1}^{c - i} p_{k,1,c - i,v - i} \right),
    \end{iarray}
    where $i \in \left[0,c \right]$ and $p_{i,1,c,v} = 0, \text{when}\;i > v$. In order to comprehend (\ref{equ:p_j_1}), we should know the number of the cases that only $i$ sub-channels with one packet exist is equal to ${\binom{c}{i}}$. In (\ref{equ:p_j_1}), assuming that there are $i$ sub-channels with one packet, ${\binom{v}{i}} \left(\frac{i}{c} \right)^{i} \left(\frac{c-i}{c} \right)^{v-i}$ is the probability that $i$ vehicles among all occupy these $i$ sub-channels with one packet and the other vehicles occupy the remainder sub-channels. $\left(\frac{1}{i} \right)^{i} i!$ is the probability to guarantee that each sub-channel in these $i$ sub-channels would be occupied by a vehicle and $\left(1 - \sum_{k=1}^{c - i} p_{k,1,c - i,v - i} \right)$ is the probability to guarantee that there is no sub-channel with one packet among the other sub-channels. In addition, the number of sub-channels with $j$ packets and its variation after one iteration is respectively denoted by $n_{j,c,v}$ and $\Delta n_{j,c',v'}$. Either of them satisfies 
    \begin{equation}\label{equ:satisfied condition 2}
    \sum_{j=0}^{v} n_{j,c,v} = c
    \end{equation}
    \begin{equation}\label{equ:satisfied condition 3}
    \sum_{j=0}^{v} j  n_{j,c,v} = v.
    \end{equation}
    
    We also define $b_{j,c',v'}$ as the expectation of $\Delta n_{j,c',v'}$. Therefore, $b_{j,c',v'}$ can be expressed as   
    \begin{equation}\label{equ:b}
    b_{j,c',v'} = \mathbb{E} \left[\Delta n_{j,c',v'} \right] = \sum_{i=0}^{c} i  p_{i,j,c',v'},
    \end{equation}    
    where $c'$ denotes the number of available sub-channels in the last iteration $k-1$ and $v'$ denotes the number of vehicles that have re-selected resources in this iteration $k$. Therefore, $c'$ can be replaced by $\left(n_{0,c,v} \right)_{k-1}$, and $v'$ can be replaced by $n_\text{rs}$.
    
    When the number of sub-channels with packet collision is equal or greater than three, that is $\sum_{j=2}^{v} n_{j,c,v} \geq 3$, the variable $n_{j,c,v}$ in each iteration can be expressed as
    \begin{iarray}\label{equ:c}
    \left(n_{0,c,v} \right)_{k} &=& \Delta n_{0,\left(n_{0,c,v} \right)_{k-1},n_\text{rs}}, \nonumber\\
    \left(n_{1,c,v} \right)_{k} &=& \left(n_{1,c,v} \right)_{k-1} + \Delta n_{1,\left(n_{0,c,v} \right)_{k-1},n_\text{rs}}, \nonumber\\
    \left(n_{2,c,v} \right)_{k} &=& \left(1 - \frac{3 \cdot 0.5}{\sum_{j=2}^{v} \left(n_{j,c,v} \right)_{k-1}} \right)  \left(n_{2,c,v} \right)_{k-1} + \Delta n_{2,\left(n_{0,c,v} \right)_{k-1},n_\text{rs}}, \nonumber\\
    \cdots \nonumber\\
    \left(n_{v,c,v} \right)_{k} &=& \left(1 - \frac{3 \cdot 0.5}{\sum_{j=2}^{v} \left(n_{j,c,v} \right)_{k-1}} \right)  \left(n_{v,c,v} \right)_{k-1} + \Delta n_{v,\left(n_{0,c,v} \right)_{k-1},n_\text{rs}},
    \end{iarray}
    where    
    \begin{equation}
    \label{equ:n_resel}
        n_\text{rs} = \frac{3 \sum_{j=2}^{v} j  n_{j,c,v}}{2 \sum_{j=2}^{v} n_{j,c,v}}.
    \end{equation}
    
    We multiply both sides of the equations (\ref{equ:c}) by their respective $j$ and accumulate them from $j=2$ to $j=v$. Then we can get    
    \begin{iarray}\label{equ:f}
    \left(\sum_{j=2}^{v} j  n_{j,c,v} \right)_{k} &=& \left(1 - \frac{3 \cdot 0.5}{\sum_{j=2}^{v} \left(n_{j,c,v} \right)_{k-1}} \right)  \left(\sum_{j=2}^{v} j  n_{j,c,v} \right)_{k-1} + \sum_{j=2}^{v} j  \Delta n_{j,\left(n_{0,c,v} \right)_{k-1},n_\text{rs}},
    \end{iarray}    
    where the left side of the equation (\ref{equ:f}) indicates the number of colliding packets. If the number of colliding packets decreases after every iteration, the convergence can be guaranteed. Equivalently, if we prove the following inequality (\ref{equ:g1}) to be satisfied, we can also prove the convergence.    
    \begin{equation}\label{equ:g1}
    \mathbb{E} \left[\frac{3}{2} \sum_{j=2}^{v} \frac{j  n_{j,c,v}}{\sum_{j=2}^{v} n_{j,c,v}} - \sum_{j=2}^{v} j  \Delta n_{j,n_{0,c,v},n_\text{rs}}\right] > 0.
    \end{equation}
    
    We find that the previous item of the inequality can be replaced by $\mathbb{E} [n_\text{rs}]$ according to (\ref{equ:n_resel}), so the inequality (\ref{equ:g1}) can be converted to    
    \begin{equation}\label{equ:i}
    \mathbb{E} \left[Z\right] = \mathbb{E} \left[n_\text{rs} - \sum_{j=2}^{v} j  \Delta n_{j,n_{0,c,v},n_\text{rs}}\right] > 0,
    \end{equation}    
    where the left side of the inequality (\ref{equ:i}) is denoted by $\mathbb{E}\left[Z\right]$. When the number of sub-channels with packet collision is less than three, that is $\sum_{j=2}^{v} n_{j,c,v} < 3$, we can get the same inequality as (\ref{equ:i}). Since $\Delta n_{j,n_{0,c,v},n_\text{rs}} = 0$, when $j > n_\text{rs}$, it converts to $n_\text{rs} - \sum_{j=2}^{n_\text{rs}} j  \Delta n_{j,n_{0,c,v},n_\text{rs}} > 0$.
    
    According to (\ref{equ:satisfied condition 3}), we can get $\sum_{j=0}^{n_\text{rs}} j  \Delta n_{j,c,n_\text{rs}} = n_\text{rs}$, which is $\sum_{j=2}^{n_\text{rs}} j  \Delta n_{j,c,n_\text{rs}} = n_\text{rs} - \Delta n_{1,n_{0,c,v},n_\text{rs}}$. Therefore,     
    \begin{iarray}\label{equ:m}
    Z &=& n_\text{rs} - \sum_{j=2}^{n_\text{rs}} j  \Delta n_{j,n_{0,c,v},n_\text{rs}} = n_\text{rs} - n_\text{rs} + \Delta n_{1,n_{0,c,v},n_\text{rs}} = \Delta n_{1,n_{0,c,v},n_\text{rs}}.
    \end{iarray}
    
\begin{figure}[t]
        \centerline{\includegraphics[width = 0.55\columnwidth ]{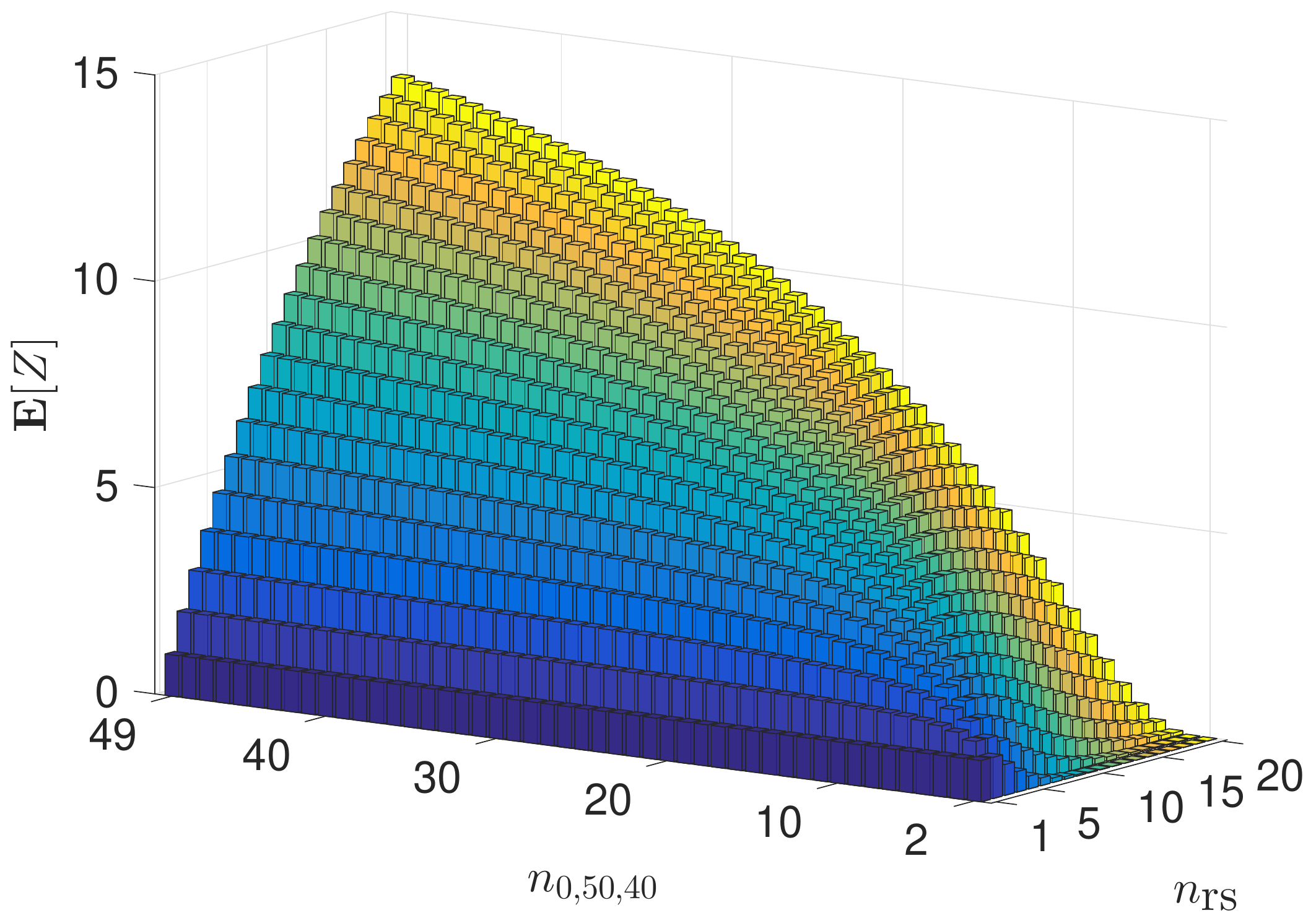}}
        \caption{Convergence analysis of collision avoidance. The figure shows $\mathbb{E} [\Delta n_{1,n_{0,c,v},n_\text{rs}}] > 0$ within all possible values, where $\min\{\mathbb{E} [Z]\} = 3.81 \cdot 10^{-5}$.}
        \label{fig:convergence_ca}
    \end{figure}

    Based on (\ref{equ:p_j_1}) and (\ref{equ:b}), we use MATLAB to calculate $b_{1,n_{0,c,v},n_\text{rs}} = \mathbb{E} \left[\Delta n_{1,n_{0,c,v},n_\text{rs}}\right]$. In practice, $v < c$, which means $n_{0,c,v} \geq 1$. When $\sum_{j=2}^{v} n_{j,c,v} > 0$, $n_{0,c,v} \geq 2$. Therefore, in our scheme, $n_{0,c,v} \in \left[2,c-1\right]$ and $n_\text{rs} \in \left[1,\left \lceil 0.5 v \right \rceil\right]$. To verify the convergence, we set $c = 50$ and $v = 40$. The result is given in Fig. \ref{fig:convergence_ca}.	
    
    The figure shows $\mathbb{E} \left[\Delta n_{1,n_{0,c,v},n_\text{rs}}\right] > 0$ within all possible values, where $\min \left\{\mathbb{E} \left[Z\right]\right\} = 3.81 \cdot 10^{-5}$, and $\Delta n_{1,n_{0,c,v},n_\text{rs}} \geq 0$, which verify the convergence.
    
    \section{Analysis of Static Vehicular Traffic Flow}
    \label{sec_static}
    In the scenario of static vehicular traffic flow, vehicles in one communication range will not leave the range and no new vehicles will enter the range, so the number of the vehicles in the range is constant as well. When the scheme is converged, the average AoI of reception of each vehicle can be expressed by
    \begin{equation}\label{equ:average_AoI_1}
        \mathbb{E}_{i} \left[a_{i}\right] = \frac{1}{v} \sum_{i=1}^{v} a_{i},
    \end{equation}
    where $v \leqslant c$ and $i$ indicates the receiving vehicle $i$. The description of the other variables used in the analysis can be found in Table \ref{tab:Variables_Static}. Since
    \begin{iarray}
        a_{i} &=& \mathbb{E}_{j} [a_{i,j}] = \frac{1}{v-1} \sum_{j=1, j\neq i}^{v} a_{i,j},\\
        a_{i,j} &=& \mathbb{E}_{t} [a_{i,j,t}] = \frac{1}{T} \sum_{t=1}^{T} a_{i,j,t},
    \end{iarray}
    where $j$ indicates the transmitting vehicle $j$ and $t$ indicates time $t$, (\ref{equ:average_AoI_1}) can be expressed by
    \begin{equation}\label{equ:E_i}
        \mathbb{E}_{i} [a_{i}] = \frac{1}{v} \sum_{i=1}^{v} \frac{1}{v-1} \sum_{j=1, j\neq i}^{v} \frac{1}{T} \sum_{t=1}^{T} a_{i,j,t},
    \end{equation}
    where
    \begin{equation}
        a_{i,j,t} = \left \{ \begin{array}{ll}
            a_{i,j,t-1} + 1, &\text{if $i$ does not receive from $j$},  \\
            0, &\text{if $i$ receives from $j$}, 
        \end{array}  \right.
    \end{equation}
    and $a_{i,j,\text{0}} = 0$.
    
    \begin{table} [t] 
    \centering 
    \caption{Variables in Performance Analysis for Static Traffic Flow} 
    \label{tab:Variables_Static}
    \footnotesize
    \begin{tabular}{c c|c c}  
    \hline
    \textbf{Variables}&\textbf{Description}&\textbf{Variables}&\textbf{Description}\\
    \hline
    $N_\text{subCH}$ &  \tabincell{c}{Number of Sub-channels \\in a Sub-frame} & $T$ & \tabincell{c}{Period where Relative Locations of the \\Sub-channels Occupied by Vehicles Keep the Same} \\
    \hline
    $\alpha_{\text{RRI}}$ &  Resource Reservation Interval & $\mathcal{J}_{i}$ & \tabincell{c}{Set of Vehicle $j$ Transmitting in \\the Same Sub-frame as Vehicle $i$} \\
    \hline
    $c$ & \tabincell{c}{Number of Sub-channels \\during an RRI} & $\mathcal{J}_{\bar{i}}$ & \tabincell{c}{Set of Vehicle $j$ Transmitting in \\Different Sub-frames from Vehicle $i$}\\
    \hline
    $a_{i}$ & Average AoI of Reception of Vehicle $i$ & $\mathcal{I}_{t}$ & Set of Vehicle $i$ Transmitting at Time $t$ \\
    \hline
    $a_{i,j}$ & \tabincell{c}{Average AoI of Reception \\of Vehicle $i$ with Vehicle $j$} & $\mathcal{I}_{\bar{t}}$ & Set of Vehicle $i$ not Transmitting at Time $t$ \\
    \hline
    $a_{i,j,t}$ & \tabincell{c}{Instantaneous AoI of Reception of \\Vehicle $i$ with Vehicle $j$ at Time $t$} & $\mathcal{J}_{\bar{i},t}$ & \tabincell{c}{Set of Vehicle $j$ Transmitting in Different Sub-frames \\from Vehicle $i$ and Transmitting at Time $t$}\\
    \hline
    $v$ & \tabincell{c}{Total Number of vehicles \\($v \leqslant c=\alpha_{\text{RRI}}  N_{\text{subCH}}$)} & $\mathcal{J}_{\bar{i},\bar{t}}$ & \tabincell{c}{Set of Vehicle $j$ Transmitting in Different Sub-frames \\from Vehicle $i$ and not Transmitting at Time $t$}\\
    \hline
    \end{tabular}  
    \end{table} 
    
    In order to further calculate $a_{i,j,t}$, six sets are defined to calculate $a_{i,j,t}$ from four parts. The six sets are expressed as $\mathcal{J}_{i}$, $\mathcal{J}_{\bar{i}}$, $\mathcal{I}_{t}$, $\mathcal{I}_{\bar{t}}$, $\mathcal{J}_{\bar{i},t}$ and $\mathcal{J}_{\bar{i},\bar{t}}$, where the description of them can be found in Table \ref{tab:Variables_Static} as well. If it is assumed that all vehicles adopt the same RRI, based on the definition, the expectation of the number of vehicles in the corresponding sets and the relation between the sets can be determined in the following equation.
    \begin{iarray}
        \mathbb{E}[|\mathcal{J}_{i}|] &=& \frac{N_\text{subCH}-1}{c-1} (v-1), \label{equ:J_i1}\\
        \mathbb{E}[|\mathcal{J}_{\bar{i}}|] &=& \frac{c-N_\text{subCH}}{c-1} (v-1),\\
        \mathbb{E}[|\mathcal{I}_{t}|] &=& \frac{1}{\alpha_{\text{RRI}}}  v,\\
        \mathbb{E}[|\mathcal{I}_{\bar{t}}|] &=& \frac{\alpha_{\text{RRI}}-1}{\alpha_{\text{RRI}}}  v,\\
        \mathbb{E}[|\mathcal{J}_{\bar{i},t}|] &=& \frac{c-N_\text{subCH}}{c-1} (v-1)  \frac{1}{\alpha_{\text{RRI}}}, \label{equ:J_i2,t1}\\
        \mathbb{E}[|\mathcal{J}_{\bar{i},\bar{t}}|] &=& \frac{c-N_\text{subCH}}{c-1} (v-1)  \frac{\alpha_{\text{RRI}}-2}{\alpha_{\text{RRI}}-1}, \label{equ:J_i2,t2}
    \end{iarray}
    where $i \in \mathcal{I}_{\bar{t}}$ in (\ref{equ:J_i2,t1}) and (\ref{equ:J_i2,t2}), $\mathcal{J}_{\bar{i},t} \cap \mathcal{J}_{\bar{i},\bar{t}} = \phi$ and $\mathcal{J}_{\bar{i},t} \cup \mathcal{J}_{\bar{i},\bar{t}} = \mathcal{J}_{\bar{i}}$. Based on the six sets, $a_{i,j,t}$ can be calculated through four parts, which can be expressed by
    \begin{equation}
        a_{i,j,t} = \left \{ \begin{array}{ll}
            a_{i,j,t-1} + 1, &j \in \mathcal{J}_{i},  \\
            a_{i,j,t-1} + 1, &j \in \mathcal{J}_{\bar{i}} \text{ and } i \in \mathcal{I}_{t},  \\
            0, &j \in \mathcal{J}_{\bar{i},t} \text{ and } i \in \mathcal{I}_{\bar{t}}, \\
            a_{i,j,t-1} + 1, &j \in \mathcal{J}_{\bar{i},\bar{t}} \text{ and } i \in \mathcal{I}_{\bar{t}}.
        \end{array}  \right.
    \end{equation}
    When $j \in \mathcal{J}_{i}$, $a_{i,j,0} = \frac{1}{\alpha_{\text{RRI}}}\left[0+1+\cdots+(\alpha_{\text{RRI}}-1)\right] = \frac{\alpha_{\text{RRI}}-1}{2}$ and there is no update during $T$. When $j \in \mathcal{J}_{\bar{i}} \text{ and } i \in \mathcal{I}_{t}$ or $j \in \mathcal{J}_{\bar{i},\bar{t}} \text{ and } i \in \mathcal{I}_{\bar{t}}$, updates occur during $T$ so that $a_{i,j,t}$ may be $1,2,\cdots, \text{ or } \alpha_{\text{RRI}}-1$. Further, when $j \in \mathcal{J}_{\bar{i},\bar{t}} \text{ and } i \in \mathcal{I}_{\bar{t}}$, the sub-frame that vehicle $i$ transmitting should be excluded. Therefore, 
    \begin{equation}
        a_{i,j,t} = \left \{ \begin{array}{ll}
            t + \frac{\alpha_{\text{RRI}}-1}{2}, &j \in \mathcal{J}_{i},  \\
            \frac{1}{\alpha_{\text{RRI}}-1} \sum_{k=1}^{\alpha_{\text{RRI}}-1} k, &j \in \mathcal{J}_{\bar{i}} \text{ and } i \in \mathcal{I}_{t},  \\
            0, &j \in \mathcal{J}_{\bar{i},t} \text{ and } i \in \mathcal{I}_{\bar{t}}, \\
            \frac{1}{\alpha_{\text{RRI}}-2} \sum_{k=1}^{\alpha_{\text{RRI}}-1} k \left(1 -\frac{1}{\alpha_{\text{RRI}}-1}\right), &j \in \mathcal{J}_{\bar{i},\bar{t}} \text{ and } i \in \mathcal{I}_{\bar{t}},
        \end{array} \right.
    \end{equation}
    and
    \begin{equation}\label{equ:A_i,j,t}
        a_{i,j,t} = \left \{ \begin{array}{ll}
            t + \frac{\alpha_{\text{RRI}}-1}{2}, &j \in \mathcal{J}_{i},  \\
            \frac{\alpha_{\text{RRI}}}{2}, &j \in \mathcal{J}_{\bar{i}} \text{ and } i \in \mathcal{I}_{t},  \\
            0, &j \in \mathcal{J}_{\bar{i},t} \text{ and } i \in \mathcal{I}_{\bar{t}}, \\
            \frac{\alpha_{\text{RRI}}}{2}, &j \in \mathcal{J}_{\bar{i},\bar{t}} \text{ and } i \in \mathcal{I}_{\bar{t}}.
        \end{array}  \right.
    \end{equation}
    
    Based on (\ref{equ:J_i1})-(\ref{equ:J_i2,t2}) and (\ref{equ:A_i,j,t}), (\ref{equ:E_i}) can be calculated in the following equation.
    \begin{iarray}\label{equ:E_i_static}
        \mathbb{E}_{i} [a_{i}] &=& \frac{1}{v} \sum_{i=1}^{v} \frac{1}{v-1} \sum_{j=1, j\neq i}^{v} \frac{1}{T} \sum_{t=1}^{T} a_{i,j,t}, \nonumber\\
        &=& \frac{1}{Tv(v-1)} \left(\sum_{t=1}^{T} \sum_{i=1}^{v} \sum_{j \in \mathcal{J}_{i}} a_{i,j,t} + \sum_{t=1}^{T} \sum_{i=1}^{v} \sum_{j \in \mathcal{J}_{\bar{i}}} a_{i,j,t}\right), \nonumber\\
        &=& \frac{1}{Tv(v-1)} \left(\sum_{t=1}^{T} \sum_{i=1}^{v} \sum_{j \in \mathcal{J}_{i}} a_{i,j,t} \right.+ \sum_{t=1}^{T} \sum_{i \in \mathcal{I}_{t}} \sum_{j \in \mathcal{J}_{\bar{i}}} a_{i,j,t} + \sum_{t=1}^{T} \sum_{i \in \mathcal{I}_{\bar{t}}} \sum_{j \in \mathcal{J}_{\bar{i},t}} a_{i,j,t}\nonumber\\
        &&\left. + \sum_{t=1}^{T} \sum_{i \in \mathcal{I}_{\bar{t}}} \sum_{j \in \mathcal{J}_{\bar{i},\bar{t}}} a_{i,j,t}\right), \nonumber\\
        &=& \frac{1}{Tv(v-1)} \left(\sum_{t=1}^{T} v  \frac{N_\text{subCH}-1}{c-1} (v-1) \right.  \left(t+\frac{\alpha_{\text{RRI}}-1}{2}\right) + \frac{Tv}{\alpha_{\text{RRI}}}  \frac{c-N_\text{subCH}}{c-1} (v-1)  \nonumber\\
        && \cdot \frac{\alpha_{\text{RRI}}}{2} + \frac{Tv(\alpha_{\text{RRI}}-1)}{\alpha_{\text{RRI}}}  \frac{c-N_\text{subCH}}{c-1}  \frac{v-1}{\alpha_{\text{RRI}}-1} \cdot 0 + \frac{Tv(\alpha_{\text{RRI}}-1)}{\alpha_{\text{RRI}}}  \frac{c-N_\text{subCH}}{c-1} (v-1) \nonumber\\
        && \left.\cdot \frac{\alpha_{\text{RRI}}-2}{\alpha_{\text{RRI}}-1}  \frac{\alpha_{\text{RRI}}}{2}\right), \nonumber\\
        &=& \frac{(N_\text{subCH}-1)(T+\alpha_{\text{RRI}})}{2(c-1)} + \frac{(c-N_\text{subCH})(\alpha_{\text{RRI}}-1)}{2(c-1)},
    \end{iarray}
    where average AoI in static scenario is irrelevant of the number of vehicles $v$ or CBR, but $v \leqslant c$ should be complied.
    
    Since $c=\alpha_{\text{RRI}}  N_{\text{subCH}}$, (\ref{equ:E_i_static}) can be easily expressed by
    \begin{equation}
        \mathbb{E}_{i} [a_{i}] = \frac{1}{2N_\text{subCH}} \left(\left(N_\text{subCH} \alpha_{\text{RRI}} - 1 \right) + \frac{N_\text{subCH} \left(N_\text{subCH} - 1\right) \left(T+1 \right)}{N_\text{subCH} \alpha_{\text{RRI}} - 1} \right) + \frac{1-N_\text{subCH}}{2N_\text{subCH}}.
    \end{equation}
    When the equation $N_\text{subCH} \alpha_{\text{RRI}} - 1 = \frac{N_\text{subCH} \left(N_\text{subCH} - 1\right) \left(T+1 \right)}{N_\text{subCH} \alpha_{\text{RRI}} - 1}$ is satisfied, the minimum average AoI can be obtained, whereby the positive solution of $\alpha_{\text{RRI}}$ of the equation is the theoretically optimal RRI. Considering that RRI is a integer and the set of the available RRI is discrete, the practically optimal RRI should be an available RRI that is close to the theoretical result.

    \section{Analysis of Dynamic Vehicular Traffic Flow}
    \label{sec_dynamic}
    In the scenario of dynamic vehicular traffic flow, it is assumed that vehicles in one communication range will arrive in and leave the range in a dynamic rate, but the number of the vehicles in the range is still constant. In the performance analysis of this part, the expectation of the arriving and leaving proportion are respectively equal to $x$ and $y$, where the description of the variables used in this section is shown in Table \ref{tab:Variables_Dynamic} as well. In order to simplify the calculation, it is assumed that 
    \begin{itemize}
        \item both $x$ and $y$ are constant and $x=y$,
        \item the minimum calculation window is an RRI,
        \item the scheme can be converged in each RRI,
        \item the packet will not collide if it has collided in the last RRI.
    \end{itemize}
    
    \begin{table} [t] 
    \centering 
    \caption{Variables in Performance Analysis for Dynamic Traffic Flow} 
    \label{tab:Variables_Dynamic}
    \footnotesize
    \begin{tabular}{c c|c c}  
    \hline
    \textbf{Variables}&\textbf{Description}&\textbf{Variables}&\textbf{Description}\\
    \hline
    $x$ & \tabincell{c}{Expectation of Proportion of the \\Number of Vehicles that Arrive \\in a Communication Range \\in One Unit of Time} & $c_{t'}$ &  \tabincell{c}{Number of Occupied Sub-channels in \\the $t'$-th RRI Window without Considering \\the Number of Arriving and Leaving Vehicles}  \\
     \hline
    $y$ & \tabincell{c}{Expectation of Proportion of the \\Number of Vehicles that Leave \\a Communication Range \\in One Unit of Time} & $b_{j,c,v}$ & \tabincell{c}{Expectation of the Number of Sub-channels that are \\Respectively Occupied by $j$ Vehicles When $v$ Vehicles \\Randomly Select $v$ Sub-channels among $c$ Sub-channels}  \\
     \hline
    $v_{\text{0}}$ & Initial Number of Vehicles & $n_{\text{col},t'}$ & Number of Collision Packets in the $t'$-th RRI Window \\
     \hline
    $t'$ &  $t'$-th RRI Window  \\
    \hline
    \end{tabular}  
    \end{table} 
    
    Based on the assumptions mentioned above, when vehicle $i$ cannot update the status of vehicle $j$ due to the collision, the AoI increment in an RRI is equal to $\alpha_{\text{RRI}}^{\text{2}}$. If there are $n_{\text{col},t'}$ packets collided in an RRI, the sum of the AoI increments equals to $n_{\text{col},t'}  \alpha_{\text{RRI}}^{\text{2}}$. Besides, only the collision packets in different sub-frames from vehicle $i$ can result in the AoI increment of vehicle $i$, so the AoI increment of vehicle $i$ can be expressed by $\frac{c-N_{\text{subCH}}}{c-1}  n_{\text{col},t'}  \alpha_{\text{RRI}}^{\text{2}}$. Based on the conclusion in the scenario of static vehicular traffic flow (\ref{equ:E_i_static}), the average AoI of reception of each vehicle in dynamic scenario can be expressed by
    \begin{iarray}\label{equ:average_AoI}
        \mathbb{E}_{i} [a_{i}] &=& \frac{(N_\text{subCH}-1)(T+\alpha_{\text{RRI}})}{2(c-1)} + \frac{(c-N_\text{subCH})(\alpha_{\text{RRI}}-1)}{2(c-1)} \nonumber\\ 
        &&+ \frac{1}{Tv'(v_{\text{0}}-1)} \sum_{t'=1}^{\frac{T}{\alpha_{\text{RRI}}}} \sum_{i=1}^{v'} \frac{c-N_{\text{subCH}}}{c-1}  n_{\text{col},t'}  \alpha_{\text{RRI}}^{\text{2}},
    \end{iarray}
    where $v_{\text{0}} \leqslant c$ and the range of $t'$ is from $1$ to $\frac{T}{\alpha_{\text{RRI}}}$ since $t'$ indicates the index of the RRI. In addition, $v'$ here indicates the number of the vehicles that keep staying in the communication range from beginning to end and $v'$ should be at least one, which means the communication range could be not only static but also mobile so that the range could follow one or several vehicles.
    
    In order to get the final result of (\ref{equ:average_AoI}), the variable $n_{\text{col},t'}$ is necessary to be calculated. Considering an RRI as a calculation window, since it is assumed that the scheme can be converged in each RRI, the number of the collision packets in the $t'$-th RRI window should be
    \begin{iarray}
        n_{\text{col},t'} &=& \left(\sum_{j=2}^{A_{\text{r}}} j  b_{j,c,A_{\text{r}}} + b_{\text{1},c,A_{\text{r}}}  \frac{c_{t'}-L_{\text{e}}}{c} + \frac{c_{t'}-L_{\text{e}}}{c} \sum_{j=1}^{A_{\text{r}}} b_{j,c,A_{\text{r}}} \right)  \frac{xv_{\text{0}}\alpha_{\text{RRI}}}{1000A_{\text{r}}},
    \end{iarray}
    where $xv_{\text{0}}$ indicates the number of vehicles that arrive in one unit of time, $yv_{\text{0}}$ indicates the number of vehicles that leave in one unit of time, $A_{\text{r}} = \left \lceil \frac{xv_{\text{0}}\alpha_{\text{RRI}}}{1000} \right \rceil$ indicates the number of arriving vehicles in each RRI and $L_{\text{e}} = \frac{yv_{\text{0}}\alpha_{\text{RRI}}}{1000}$ indicates the number of leaving vehicles in each RRI. As assumed that the scheme can be converged in each RRI, so $c_{t'} = c_{t'-1} + (x-y)v_{\text{0}} = c_{t'-\text{1}}$. And because $c_{\text{0}} = v_{\text{0}}$, $c_{t'} = v_{\text{0}}$. Besides, according to (\ref{equ:satisfied condition 2}) and (\ref{equ:satisfied condition 3}), $\sum_{j=2}^{A_{\text{r}}} j  b_{j,c,A_{\text{r}}} = A_{\text{r}} - b_{\text{1},c,A_{\text{r}}}$ and $\sum_{j=1}^{A_{\text{r}}} b_{j,c,A_{\text{r}}} = c-b_{\text{0},c,A_{\text{r}}}$, therefore
    \begin{iarray}\label{equ:n_col}
        n_{\text{col},t'} &=& \left(A_{\text{r}} - b_{\text{1},c,A_{\text{r}}} + b_{\text{1},c,A_{\text{r}}}  \frac{v_{\text{0}}-L_{\text{e}}}{c} + \frac{v_{\text{0}}-L_{\text{e}}}{c} (c-b_{\text{0},c,A_{\text{r}}}) \right)  \frac{xv_{\text{0}}\alpha_{\text{RRI}}}{1000A_{\text{r}}}=n_{\text{col}},
    \end{iarray}
    where $b_{\text{1},c,A_{\text{r}}}$ and $b_{\text{0},c,A_{\text{r}}}$ can be calculated by (\ref{equ:b}) and $p_{i,0,c,v} = \binom{c}{i}  \left(1-\frac{i}{c}\right)^{v}  \left(1 - \sum_{k=1}^{c-i} p_{k,\text{0},c-i,v}\right)$.
    
    Based on (\ref{equ:average_AoI}) and (\ref{equ:n_col}), the average AoI of reception of each vehicle in dynamic scenario can be further calculated, where
    \begin{iarray}\label{equ:average_AoI_final}
        \mathbb{E}_{i} [a_{i}] &=& \frac{(N_\text{subCH}-1)(T+\alpha_{\text{RRI}})}{2(c-1)} + \frac{(c-N_\text{subCH})(\alpha_{\text{RRI}}-1)}{2(c-1)} \nonumber\\
        && + \frac{1}{Tv'(v_{\text{0}}-1)}  \frac{T}{\alpha_{\text{RRI}}}  v'  \frac{c-N_{\text{subCH}}}{c-1}  n_{\text{col}}  \alpha_{\text{RRI}}^{\text{2}}, \nonumber\\
        &=& \frac{(N_\text{subCH}-1)(T+\alpha_{\text{RRI}})}{2(c-1)} + \frac{(c-N_\text{subCH})(\alpha_{\text{RRI}}-1)}{2(c-1)} + \frac{(c-N_{\text{subCH}})\alpha_{\text{RRI}}}{(v_{\text{0}}-1)(c-1)}  n_{\text{col}},
    \end{iarray}
    where only the dynamic part of average AoI is related to the number of vehicles $v$ and $v \leqslant c$ should be complied as well.

    \end{document}